\documentclass[aps,prl,twocolumn,showpacs,superscriptaddress]{revtex4-1}  
\usepackage{amsmath}
\usepackage{amssymb}   
\usepackage{amsthm}   
\usepackage{bm}
\usepackage{float}
\usepackage{url}
\usepackage{graphicx}
\usepackage{algorithm,algcompatible}
\usepackage{hyperref}
\hypersetup{
  colorlinks   = true, 
  urlcolor     = blue, 
  linkcolor    = black, 
  citecolor   = black 
}

\newcommand{\logfn}[2]{\left({#1} - {#2}\right) \log \left( \frac{ {#1}}{{#2}} \right)}
\renewcommand{\d}{\text{d}}
\renewcommand{\div}[2]{\frac{\d {#1}}{\d {#2}}}
\renewcommand{\v}{\boldsymbol}

\newcommand{\be}{\begin{equation}}
\newcommand{\ee}{\end{equation}}

\newtheorem*{theorem*}{Theorem}

\usepackage{color}

\begin{document}
\raggedbottom
\title{Improved bounds on entropy production in living systems}

\author{Dominic J. Skinner} 
\affiliation{Department of Mathematics, Massachusetts Institute of Technology, Cambridge Massachusetts 02139-4307, USA}
\author{J\"{o}rn Dunkel}
\affiliation{Department of Mathematics, Massachusetts Institute of Technology, Cambridge Massachusetts 02139-4307, USA}
\date{\today}

\begin{abstract}
Living systems maintain or increase local order by working against the Second Law of Thermodynamics.
Thermodynamic consistency is restored as they dissipate heat, thereby increasing the net entropy of their environment.
Recently introduced estimators for the entropy production rate have provided major insights into the thermal efficiency of important cellular 
processes. In experiments, however, many degrees of freedom typically remain hidden to the observer, and in these cases, existing methods are not optimal.
Here, by reformulating the problem within an optimization framework, we are 
able to infer improved bounds on the rate of entropy production from partial measurements of biological systems.
Our approach yields provably optimal estimates given certain measurable transition statistics.
In particular, it can reveal non-zero heat production rates  even when non-equilibrium processes appear time symmetric 
and so may pretend to obey detailed balance.  We demonstrate the broad applicability of this framework 
by providing improved bounds on the entropy production rate in a diverse range of biological systems including 
bacterial flagella motors, growing microtubules, and calcium oscillations within human embryonic kidney cells.

\end{abstract}

\maketitle

Thermodynamic laws place fundamental limits on the efficiency and fitness of living systems~\cite{Gnesotto_2018,schrodinger_penrose_2012}. To maintain cellular order and perform essential biological functions  such as  sensing~\cite{PhysRevLett.105.048104,Mehtai_2012,Lan_2016,Nadrowski2004}, signaling~\cite{Clapham2007}, replication~\cite{Maitra406,England_2013} or locomotion~\cite{Nirody_2017},  organisms consume energy and dissipate heat.  In doing so, they increase the entropy of their environment~\cite{schrodinger_penrose_2012}, in agreement with the second law of thermodynamics~\cite{Bryant3478}.  Obtaining reliable estimates for the entropy production in living matter holds the key to understanding the physical boundaries~\cite{Crooks1999,Seifert_2016,Pietzonka_2016} that constrain the range of theoretically and practically possible biological processes~\cite{PhysRevLett.105.048104}. Recent experimental~\cite{Fodor_2016,Nadrowski2004,Rodenfels_2019}  and theoretical~\cite{Seifert_2005,Gingrich_2017,Gingrich_natPhys,Talkner_2020}  advances in the imaging and modeling of cellular and subcellular dynamics have provided groundbreaking insights into the thermodynamic efficiency of molecular motors~\cite{Pietzonka_2016,Hanggi_2009}, biochemical signaling~\cite{Bialek_2005,Horowitz_clock,Rodenfels_2019} and reaction~\cite{Rao_2016} networks, and replication~\cite{England_2013} and adaption~\cite{Lan2012} phenomena. Despite such major progress, however, it also known that the currently available entropy production estimators~\cite{Li_2019,Seifert_AnnRev} can fail under experimentally relevant conditions, especially when only a small set of observables is experimentally accessible or non-equilibrium transport currents~\cite{Esposito_coarse,Horowitz_nocurrent,Parrondo_2010} vanish. 
\par
To help overcome these limitations, we introduce here a generic optimization framework that can produce significantly improved bounds on the entropy production in living systems. We will  prove that these bounds are optimal given certain measurable statistics.  From a practical perspective, our method only requires observations of a few coarse-grained state variables of an otherwise hidden Markovian network.  We demonstrate the practical usefulness by determining improved entropy production bounds for bacterial flagella motors~\cite{Nirody_2017,BerryScience}, growing microtubules~\cite{Mitchison1984,Lacroix2014} and calcium oscillations~\cite{Clapham2007,Thurley2014} in human embryonic kidney cells.

\par 
Generally, entropy production rates can be estimated from the time series of stochastic obervables~\cite{Seifert_2012}. Thermal equilibrium systems obey the principle of detailed balance, which means that every forward trajectory is as likely to be observed as its time reversed counterpart, neutralizing the arrow of time~\cite{Parrondo_2009}. 
By contrast, living organisms operate far from equilibrium, which means that the balance between forward and reversed trajectories is broken and net fluxes may  
arise~\cite{Maes_2002,Gnesotto_2018,FakhiriPRL,Fakhri2016}.  When all microscopic details of a non-equilibrium  system are known, one can measure the rate of entropy production  by comparing the likelihoods of forward and reversed trajectories in sufficiently large data samples~\cite{Seifert_2012,Parrondo_2009}. However, in most if not all biophysical experiments, many degrees of freedom remain hidden to the observer, demanding methods~\cite{Esposito_coarse,Bisker_2017,Paulsson2009Nature} that do not require complete knowledge
of the system. A powerful alternative is provided by thermodynamic uncertainty relations (TUR) which use the and variance
of steady state currents to bound entropy production rates~\cite{Gingrich_natPhys,Li_2019,Gingrich_2017,Seifert_prl_2015,Gingrich_prl_2016,BrownianMovies,Otsubo_2020,Hawoong_2020,Hasegawa_2020,Campisi_2020}.
Although highly useful when currents can be measured~\cite{BrownianMovies,Hawoong_2020,Hasegawa_2020,Otsubo_2020}, or when the system can be externally manipulated~\cite{Polettini_2017,Bisker_2017}, these methods give, by construction,  trivial zero-bounds for current-free non-equilibrium system, such as driven one-dimensional (1D) non-periodic oscillators.  In the absence of currents,  
potential asymmetries in the forward and reverse trajectories can still be exploited to bound the entropy production rate~\cite{Horowitz_nocurrent,Roldan_2018,Parrondo_2010}, but to our knowledge, 
no existing method is capable of producing non-zero bounds when forward and reverse trajectories are statistically identical.
Moreover, even though previously bounds can become tight in some  cases~\cite{shortExp}, optimal entropy production estimators for non-equilibrium systems are in general unknown.

\par
To obtain bounds that are provably optimal under reasonable conditions on the available data, we reformulate the problem here within an optimization framework. 
Formally, given any steady-state Markovian dynamics for which only coarse-grained variables  are observable, we search over all possible Markovian systems to 
identify the one which minimizes entropy production rate while obeying the observed statistics. More specifically, our algorithmic implementation  leverages information about successive transitions, allowing us to discover non-zero bounds on entropy production even
when the coarse-grained statistics  appear time symmetric. We demonstrate this for both synthetic test data and experimental data~\cite{Nirody_2019} for flagella motors.  Subsequently, we consider  the entropy production of microtubules~\cite{Lacroix2014}, which slowly grow before rapidly shrinking in 
steady state, to show how refined coarse-graining in space and time leads to improved bounds. The final application to calcium oscillations in human embryonic kidney cells~\cite{Thurley2014} illustrates how external stimulation with drugs can increase entropy production.

\section{Results}

\begin{figure}[t!]
\includegraphics[width=0.5\textwidth]{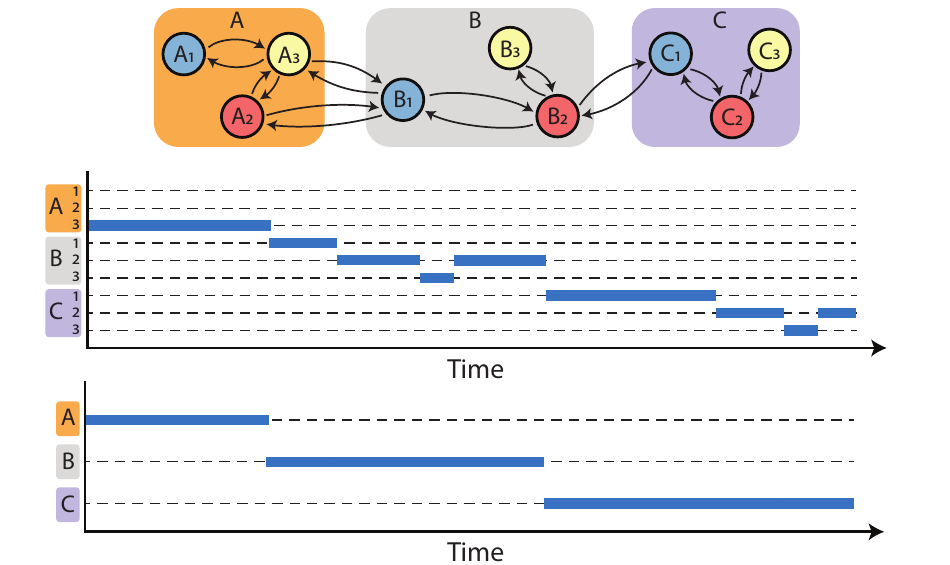}
\caption{\label{Fig:Explain}
Illustration of an underlying Markovian transition network and the coarse-grained observed system. 
The microscopic Markovian system (top) contains 9 states labeled $A_1,\dots C_3$, while the observer can only distinguish the  coarse-grained macro-states $A,B,C$ and transitions between them. 
Sample trajectory on microstates of the system (middle), and the observed macro-state trajectory (bottom), which in general exhibits non-Markovian transition dynamics.}
\end{figure}

\textbf{Theoretical background.}
Due to the large number of particles involved, classical thermodynamics can reasonably treat macroscopic processes,
like a combustion cycle in an engine, as deterministic. By contrast, theoretical descriptions of microbiological processes such as intracellular stochastic reactions~\cite{Paulsson2004,Rao_2016}, cellular sensing~\cite{Mehtai_2012,Lan2012,Wingreen_2013} and
DNA transcription and repair~\cite{Linn_2004,Hopfield_1974} must account for fluctuations~\cite{Gnesotto_2018,Jarzynski_2011}.
The most widely used framework~\cite{Seifert_2005}  for this purpose are probabilistic Markov models that assume stochastic transitions between a discrete number of states~\cite{van1992stochastic} (Fig.~\ref{Fig:Explain}). 
Here, we will merely assume that at some fundamental level such a discrete Markovian description is possible; the results below apply to all non-equilibrium processes that can be described in this manner as well as to continuous Langevin-type models that can be arbitrarily well approximated by a discrete Markovian system~\cite{SM,Kloeden1992}. 

\textbf{Exact entropy production rate.}
Our goal is to construct an estimator that comes as close as possible to the true entropy production rate $\sigma$ of the underlying microscopic Markov model, which remains hidden us. The only assumptions we shall make is that microscopic state network is connected,  that transitions between states are reversible as required by thermodynamics~\cite{Gingrich_natPhys},  and that there is no external time-dependent driving,  so that the microscopic  system~$\mathcal{S}$ will reach a unique steady state in which it spends a fraction $\pi_i$ of the time in state $i$. In this case, the true rate of entropy production $\sigma(\mathcal{S})$ is formally given by~\cite{Gingrich_natPhys}
\be
\sigma = \frac{k_B}{2} \sum_{i\neq j} (\pi_i q_{ij} - \pi_j q_{ji}) \log \left( \frac{\pi_{i}q_{ij}}{\pi_j q_{ji}} \right),
\label{e:exact_sigma}
\ee
where $q_{ij}$ is the rate at which the system transitions from microstate $i$  to microstate $j$~\cite{Gingrich_natPhys}. In principle, given all the states and a sufficiently long system trajectory, we could deduce the values of $\pi_i$ and $q_{ij}$, and hence calculate $\sigma$.  In practice, however, one typically cannot know or observe all the states, and experimental time-series measurements are only possible for severely coarse-grained macroscopic observables (Fig.~\ref{Fig:Explain}). The challenge is then to estimate $\sigma$ from such coarse-grained data.

\begin{figure*}[t!]
\includegraphics[width=\textwidth]{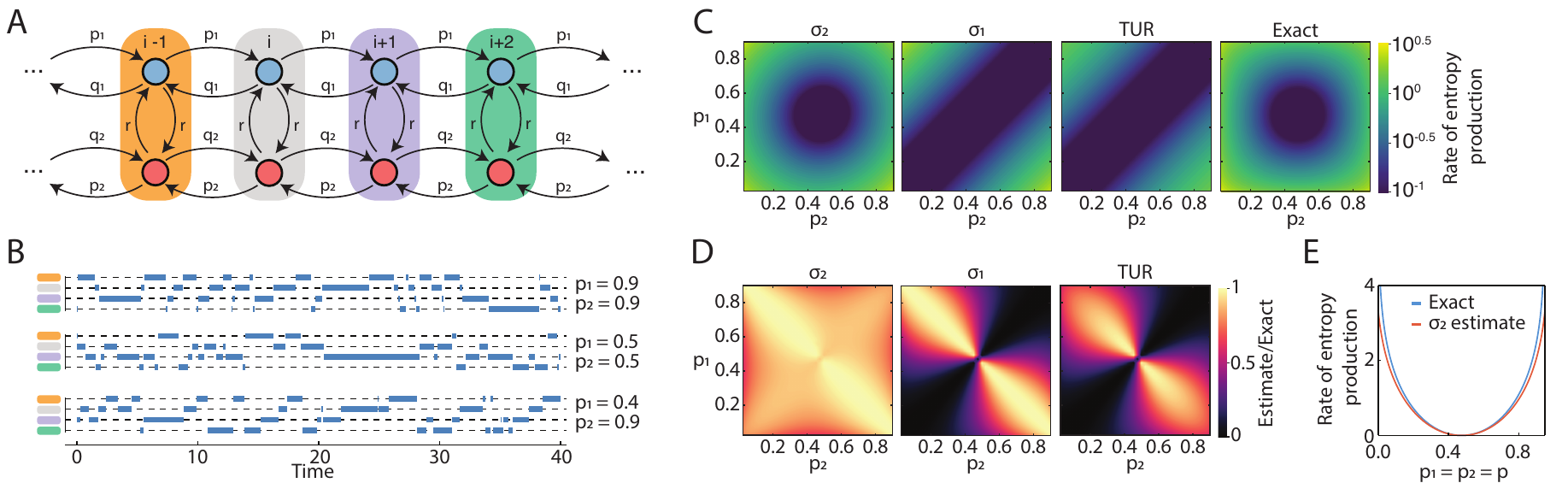}
\caption{\label{Fig:BiasedWalker}
The two-step estimator $\sigma_2$ outperforms other estimators for the switching 
biased random walk. 
(A)~Diagram for the underlying Markov dynamics, with circles representing internal states, and shaded
regions representing the observed macrostates, $i,\dots, i+3$. All waiting time distributions are exponential
with mean $\lambda = 1$, implying that $1 = r + p_1 + q_1 = r + p_2 + q_2$. Throughout this 
figure, we fix $r=0.05$, so that specifying $p_1$, and $p_2$ is sufficient to describe the system. 
(B)~Sample trajectories for different values of $p_1$ and $p_2$ in a periodic network
with 4 observed states. 
(C)~$\sigma_2$, $\sigma_1$, and TUR estimates versus the exact entropy production rates in the $(p_1,p_2)$ plane. 
(D)~The ratios between estimates and true values show that 
the $\sigma_2$ estimator provides a close fit for all values of $p_1$ and $p_2$, whereas the other estimators only 
perform well for certain combinations of transition rates. 
(E)~In the time-symmetric case $p_1 = p_2 = p$, the $\sigma_2$ estimates 
closely bound the exact values, whereas the other estimators give trivial zero-bounds. }
\end{figure*}

\textbf{Coarse-grained observables can be non-Markovian.}
Despite the Markovian nature of the underlying microscopic process, the observed coarsed grained trajectories need not be
Markovian, $2^{nd}$ order Markovian or even~$N^{th}$ order Markovian for any~$N$~\cite{Horowitz_nocurrent,SM}.
The set of macroscopic observables, $\mathcal{O}(\mathcal{S})$, therefore contains infinitely many measurements. For instance,
for the example process in Fig.~\ref{Fig:Explain}, one could measure $\hat{\pi}_A$, the fraction of the time spent in
macrostate $A$, or $\hat{\pi}_A \hat{q}_{AB}$, the rate at which $A \to B$ transitions are observed.
One could also measure more complex quantities, like $\hat{\pi}_A\hat{q}_{ABC\cdots A}$,
the rate at which trajectories are observed to take the arbitrarily long
path $ABC\cdots A$; such observables do not necessarily follow from simpler statistics.

\textbf{Bounding entropy production by solving a minimization problem.}
To reformulate the estimation of $\sigma(\mathcal{S})$ as a tractable optimization problem, let us first suppose that we are given all quantities in $\mathcal{O}(\mathcal{S})$. 
In this case, we know that the true entropy production rate of the system $\mathcal{S}$ is at least as large as the minimum entropy
production of all systems $\mathcal{R}$  with the same observed statistics
\be
\sigma(\mathcal{S}) \geq \min \left\{ \sigma (\mathcal{R}) | \mathcal{O}(\mathcal{R}) = \mathcal{O}(\mathcal{S})
\right\}.
\label{e:O}
\ee
In particular, this bound is the best possible bound without knowing further details of the underlying network 
topology, and hence the best possible estimator. 
\par
In practice, it is only feasible to measure a select few 
quantities in $\mathcal{O}$, but from these, we can build an similar estimator. Specifically, given a 
set $\mathcal{O}_k$, containing a subset of the total observables $\mathcal{O}$, we still have that
\be
\sigma(\mathcal{S}) \geq \min \left\{ \sigma (\mathcal{R}) | \mathcal{O}_k(\mathcal{R}) = 
\mathcal{O}_k(\mathcal{S}) \right\},
\label{e:O_k}
\ee
where the new estimator on the rhs. is the optimal bound given this smaller set of observables. Note that fewer 
observables provide fewer restrictions on the set of possible microsystems $\mathcal{R}$, meaning that the bound 
in Eq.~\eqref{e:O_k} is lower than that in Eq.~\eqref{e:O}.

\textbf{One-step estimator.}
A simple useful observable subset is $\mathcal{O}_1 = \{ \hat{\pi}_I \hat{q}_{IJ} \}$, containing the rates 
at which transitions $I\to J$  happen for all pairs of observed macrostates $(I,J)$. 
For the specific network topology in Fig.~\ref{Fig:Explain}, the observed statistic 
$\hat{\pi}_A \hat{q}_{AB} \in \mathcal{O}_1$,
simply counts the rate at which $A\to B$ transitions are observed, and can be expressed in terms of the 
microstates as $\hat{\pi}_A \hat{q}_{AB} = \pi_{A_2} q_{A_2 B_1} + \pi_{A_3} q_{A_3 B_1}$. Despite having to minimize over infinitely
many network topologies $\mathcal{R}$ consistent with the $\mathcal{O}_1$ statistics, finding the corresponding estimator $\sigma_1$ 
is straightforward.  This is due to the fact that,  given any network topology consistent with $\mathcal{O}_1$, one can combine two microstates in
the same macrostate in such a way that one preserves the $\mathcal{O}_1$ statistics, whilst lowering the entropy
production rate (SI~\cite{SM}). By repeatedly applying this procedure, the resulting system has no hidden states, 
every macrostate corresponds to exactly
one microstate, and the entropy production rate of this system can therefore be calculated directly.
The estimator $\sigma_1$ coincides with known estimators, $\dot{S}_\text{aff}$ in Ref.~\cite{Horowitz_nocurrent} and 
relative entropy of 2-strings in Ref.~\cite{Parrondo_2010}, but was not previously treated within an optimization
framework. However, it turns out that substantially improved entropy production bounds can be obtained by combining information from 
two successive transition steps (Fig.~\ref{Fig:BiasedWalker}).

\begin{figure*}[t]
\includegraphics[width=\textwidth]{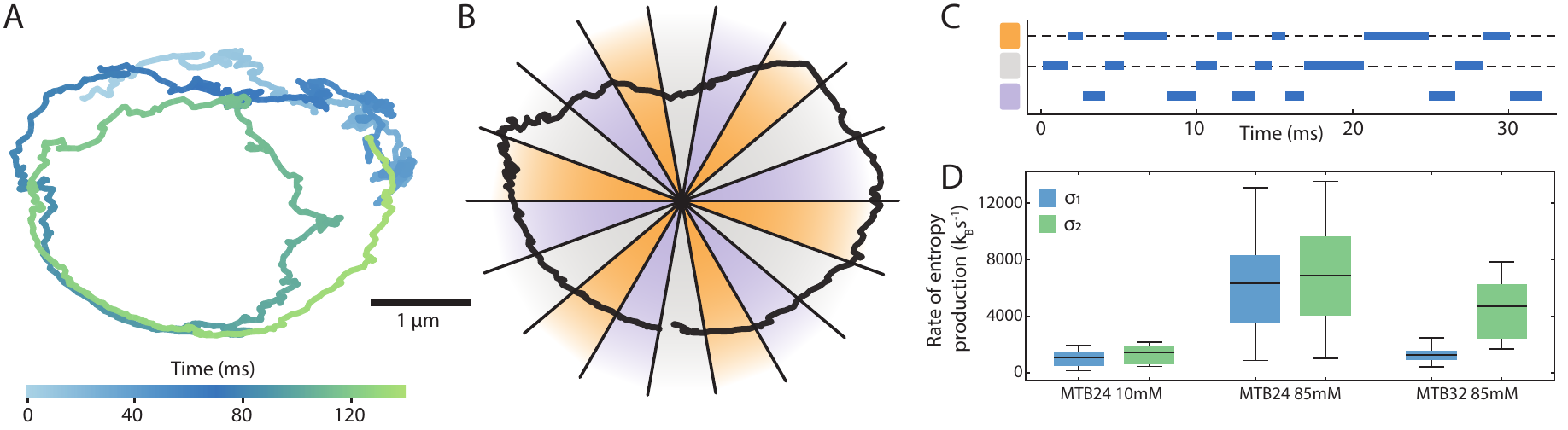}
\caption{\label{Fig:Bacteria}The estimator $\sigma_2$ bounds the rate of entropy production for the bacterial 
flagellar motor. (A) Short trajectory of bead attached to MTB32 \emph{E. coli} bacteria flagella. The bead
begins rotating clockwise, but changes direction after around 30ms, subsequently rotating counter clockwise.
(B) The plane is discretized into 3 regions (purple, orange, grey), each region is made of $N$ segments, 
here $N=6$. A short bead trajectory is overlaid.
(C) The trajectory from (B) after coarse graining onto the 3 macrostates. (D) Box plot of entropy production 
rates for different strains, sodium concentrations and estimators. The $\sigma_1$ estimator measures a similar
entropy production rate for MTB24 $10mM$ and MTB32 $85mM$, whereas the $\sigma_2$ estimator can distinguish
them.
}
\end{figure*}

\par
\textbf{Two-step estimator.}
To go beyond $\mathcal{O}_1$ statistics, we consider the set $\mathcal{O}_2 = \mathcal{O}_1 \cup \{ \hat{\pi}_I
\hat{q}_{IJK} \}$, containing the rates at which two successive transitions  $I \to J \to K$ occur for all triplets $(I,J,K)$. Knowledge of $\mathcal{O}_2$ imposes stronger constraints 
on the set of underlying Markov processes $\mathcal{R}$, promising a better bound on the entropy production rate.
In practice, performing a direct numerical minimization to obtain the corresponding estimator $\sigma_2$ is not possible due to the arbitrary complexity of permissible Markovian 
network topologies~$\mathcal{R}$. However, two exact analytic results, proved in the SI~\cite{SM}, enable us to find the best possible bound
for the combined entropy production across all edges connected to a state $J$, whilst preserving
the $\mathcal{O}_2$ statistics $\pi_I q_{IJ}$, $\pi_J q_{JI}$, $\pi_I q_{IJK}$ for any distinct 
neighboring macrostates $I,K$. Specifically, our first result enables us to
take any network $\mathcal{R}$ consistent with $\mathcal{O}_2$ and simplify its internal topology so that only $J$ has hidden states,
and further, that $J$ has no internal connections.  We show (SI~\cite{SM}) that one can always construct the simplified network in such a way that 
the entropy production rate  is lowered while remaining consistent with the $\mathcal{O}_2$ 
statistics involving $J$. Our second result proves that minimizing over this simplified topology, with arbitrarily many internal states of $J$, yields the same bound as minimizing over a system with 6 internal states for each pair of neighboring macrostates $(I,K)$. This fact makes the problem numerically tractable~\cite{liberzon2011calculus,GlobalSearch,SM}. By bounding the entropy production rate across connecting edges 
for every macrostate in this manner, we get a $\sigma_2$-bound for the total entropy production, This new estimator satisfies the 
hierarchy  $\sigma \geq \sigma_2 \geq \sigma_1$, and can be computed by observing the states visited by a suitably long 
trajectory without measuring conditional waiting time distributions~\cite{Horowitz_nocurrent,SM}.

\par
\textbf{Bounding entropy production for time symmetric observables.}
We demonstrate the performance of $\sigma_2$ relative to other estimators for a physically and biologically relevant test process, 
corresponding to a biased random walk that switches with rate $r$ between two modes of bias (Fig.~\ref{Fig:BiasedWalker}). This process represents a minimal   
model for the discretized angular dynamics of bacterial motors that switches rotation direction~\cite{Wang_2014,BerryScience}. It can also describe a particle subjected to a flashing force~\cite{Hwang_2019} or, more generally,  active Brownian on a lattice~\cite{Mandal_2018}. As shown in Fig.~\ref{Fig:BiasedWalker}A, when fixing the internal 
transition rate $r$ and assuming an exponential waiting time distributions on all internal states, the model dynamics is controlled by the two transition rate parameters $p_1$ and $p_2$,  describing right and left jumps, respectively. Coarse-grained sample trajectories, corresponding to observations of four macrostates for different combinations of $p_1$ and $p_2$, are shown in  Fig.~\ref{Fig:BiasedWalker}B. In the special case $p_1 = 1 - r - p_2$, we recover a biased random walk, or Brownian clock if made periodic~\cite{Seifert_2016}, which has effectively no hidden states. 
Alternatively, if $p_1 = p_2=p$, the observed system -- despite being out of equilibrium for all but one value of $p$ -- is completely time symmetric with every forward
path as likely to appear as every reverse path, implying vanishing net fluxes.
\par
To illustrate the benefits of leveraging multi-step information, we compare $\sigma_2$ to the one-step estimator $\sigma_1$ and also with entropy production rate  
estimates from the thermodynamic uncertainty relation (TUR)~\cite{Gingrich_natPhys,Li_2019,Gingrich_2017} (SI~\cite{SM}). We find that in the strong-flux regime, when  $p_1$ and $p_2$ are sufficiently different,  all estimators reasonably bound the true entropy production rate $\sigma$ from Eq.~\eqref{e:exact_sigma} (Fig.~\ref{Fig:BiasedWalker}C,D).
However,  as $p_1$ and $p_2$ approach each other and the net flux becomes weaker, only $\sigma_2$ gives an accurate bound (Fig.~\ref{Fig:BiasedWalker}D).  In particular, when $p_1 = p_2=p$,  the forward and reverse observables are time symmetric, so the relative entropy between them is zero~\cite{Horowitz_nocurrent}. Therefore, neither $\sigma_1$, which here coincides with the estimator, $\dot{S}_{KLD}$, in Ref.~\cite{Horowitz_nocurrent}, nor TUR can yield a non-trivial (non-zero) bound, whereas $\sigma_2$ can be computed analytically in this case (SI~\cite{SM}) and approximates the exact rate $\sigma$ well for all values of $p$ (Fig.~\ref{Fig:BiasedWalker}E). We next apply the two-step estimator $\sigma_2$ to data from recent experiments.

\begin{figure*}[ht!]
\includegraphics[width=\textwidth]{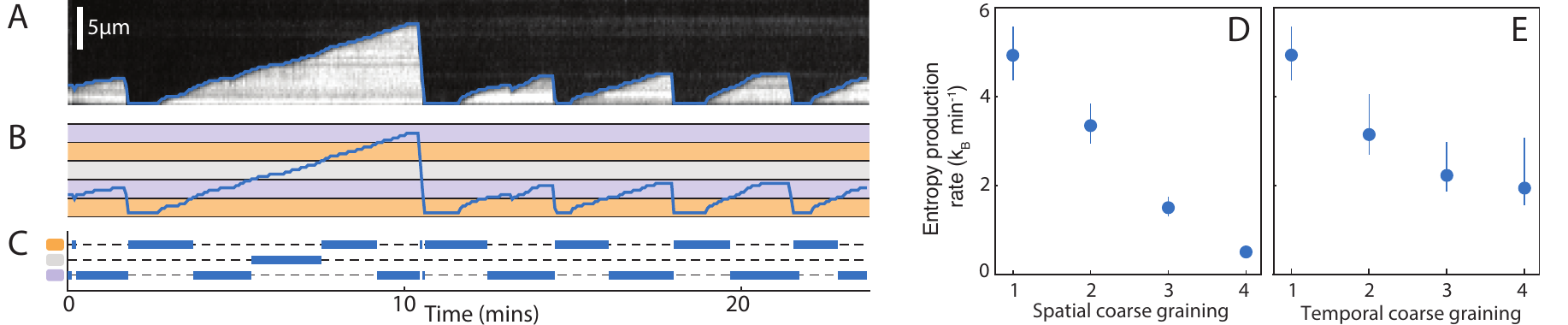}
\caption{\label{Fig:MT}Bounding the entropy production rate of steady state microtubule growth. (A) Typical kymograph of growing
microtubule showing steady growth, rapid shrinkage until vanishing, and then regrowth after nucleation. Overlaid is the segmented
length (blue line). (B) The length trajectory is discretized into 3 regions which are layered periodically. (C) The corresponding 
discretized trajectory for the system in (B). (D) The largest rate of entropy production is calculated when the regions 
are 1 pixel tall, and the inferred entropy production decreases for regions of size 2,3,4 pixels. Information is lost due to
the finite spatial resolution. Errorbars are bootstrapped $95\%$ confidence intervals. 
(E) Performing a coarse graining in time, taking every $2^{nd}$, $3^{rd}$, or every $4^{th}$ 
image, also lowers the inferred entropy production rate, but not as strongly as the spatial coarse graining.}
\end{figure*}

\textbf{Switching trajectories of bacterial flagellar motor.}
By rotating helical flagella, many species of bacteria can swim, reaching speeds of tens of body lengths per second~\cite{Nirody_2017,sowa_berry_2008}. Each flagellum is driven by a remarkable nanoscale motor, powered by a flux of ions across cytoplasmic 
membrane, which can achieve over 100 rotations per second~\cite{sowa_berry_2008,BerryScience,Berg2003}. 
Measuring the entropy production of the motor promises insights into the efficiency of small self-assembled engines and microbial locomotion~\cite{Nirody_2017,Ekeh_2020}. 
Direct experimental observations of the motor dynamics have become possible by tethering the cell, attaching a bead to the 
flagellum, and tracking the bead trajectory through high resolution microscopy~\cite{BerryScience,Nord_2017,
KRASNOPEEVA2019,Nirody_2019}. The motor-and-bead system operates in a heat bath at finite temperature, and the observed bead trajectories can be described by a Markovian Langevin-type dynamics~\cite{Nirody_2017}.  We can apply the estimators $\sigma_1,\sigma_2$ directly to measured trajectories to 
bound the entropy production rate of the motor~(SI~\cite{SM}).

\par
A representative bead trajectory for an $\emph{Escherichia coli}$ bacterium, from a recent experiment by 
Nirody \emph{et al.}~\cite{Nirody_2019}, is shown in Fig.~\ref{Fig:Bacteria}A. Measured trajectories typically follow approximately 
circular curves in the projection plane, but certain strains will stochastically
switch their rotation direction~\cite{BerryScience}. This means that, although taking place far-from-equilibrium, the process
may not obviously violate time irreversibility, limiting the applicability of previous entropy production estimators.  Bead trajectories provide a coarse-grained view of the motor system -- 
our framework allows us to coarse-grain further, dividing the total system radially 
into 3 macrostates (Fig.~\ref{Fig:Bacteria}B).  An accordingly discretized 
trajectory is shown in Fig.~\ref{Fig:Bacteria}C. From a practically perspective, having a smaller number of states can be preferable for acquiring precise transition
statistics, especially if data is limited.
\par
We estimated entropy production bounds for two sodium powered strains of $\emph{E. coli}$, comparing the non-switching strain MTB24 at  fuel concentrations of $10mM$  $Na^{+}$ and $85mM$  $Na^{+}$ with the switching strain MTB32 at $85mM$ $Na^{+}$ (Materials and Methods). For the non-switching MTB24 strain, which strongly breaks time-reversal symmetry,  the $\sigma_2$ bound does not improve significantly on the $\sigma_1$ estimate (Fig.~\ref{Fig:Bacteria}D). As expected, both estimators find that a higher ion concentration increases the bound on the entropy production rate for the non-switching MTB24 strain, as higher frequency rotations were observed. However, for the switching MTB32 strain, we find that the $\sigma_1$
significantly underestimates the entropy production rate relative to  $\sigma_2$. More specifically, the mean entropy production rate of MTB24 at low fuel concentration $10mM$  $Na^{+}$ and MTB32 at high fuel concentration $85mM$ $Na^{+}$  cannot be statistically distinguished under the
$\sigma_1$ estimator ($P<0.05$; Fig.~\ref{Fig:Bacteria}D). By contrast, the $\sigma_2$ estimator clearly distinguishes  ($P<0.01$) between the two experiments, yielding high-fuel entropy production estimates that are consistent for both strains (Fig.~\ref{Fig:Bacteria}D). Corroborating the results from biased random walk test case, this application highlights the importance of incorporating multi-transition information when estimating entropy production for non-equilibrium system with small net fluxes.
\par
From a broader conceptual perspective, it is worth emphasizing that the entropy production bounds were obtained without assuming any particular model for the motor's dynamics,
precise measurements of ion concentrations, or a rheological characterization of the medium. The rate estimates can be used to gain insights into the working principles and fuel consumption of bacterial motors. For example, measurements of the ion motive force suggest that the free energy change of a single ion transit is around 
$6k_BT$~\cite{sowa_berry_2008}. Combining this with the estimates in Fig.~\ref{Fig:Bacteria}D, we can bound the average rate of ion consumption as $\geq 1,000 s^{-1}$ for
the $85mM$ fuel concentrations.

\begin{figure*}[t!]
\includegraphics[width=\textwidth]{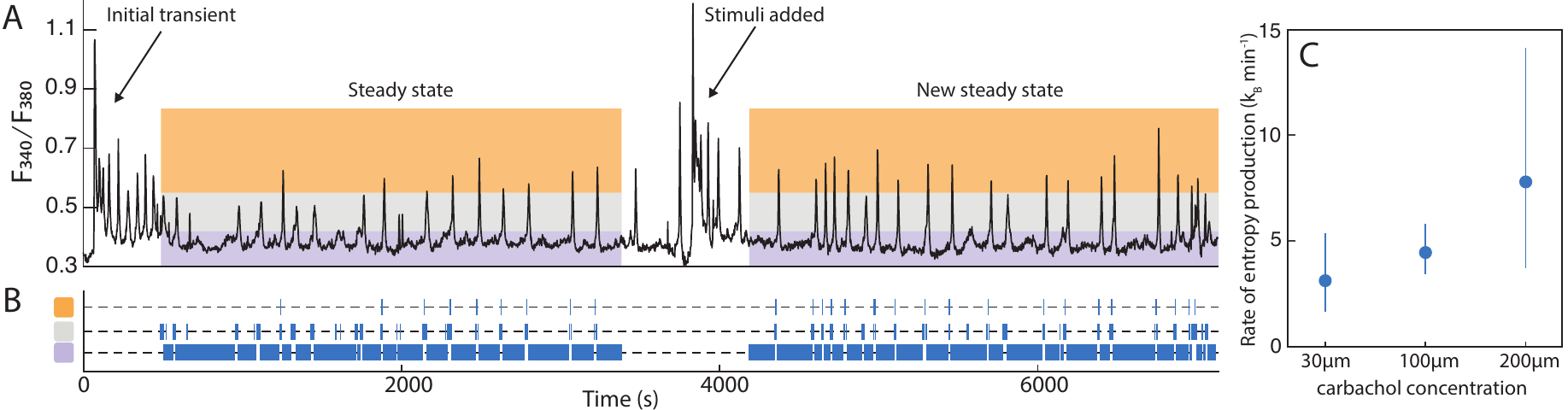}
\caption{\label{Fig:HEKOsc}Bounding the entropy production rate of calcium oscillations within human kidney cells. (A) 
Ratio of fluorecence at different wavelengths, as a proxy for calcium concentration, against time. Initially cells
are exposed to $30\mu M$ carbachol and reach a statistically steady state. After around an hour they are exposed to
a higher level of carbachol, and reach a new steady state. During these steady states, we
partition space into 3 regions as shown. (B) Discretized trajectory on the 3 regions. (C) Bounds computed after the initial $30\mu M$ 
stimulation, and after the subsequent re-stimulation with $100 \mu M$ or $200\mu M$ carbachol, reveal a substantial increase in entropy production in response to the 
 second stimulus. Errorbars are bootstrapped $95\%$
confidence intervals.}
\end{figure*}
\textbf{Dynamic instability of microtubule growth.}
Microtubules are a core component of eukaryotic cells, providing structural stability, enabling intracellular transport,
and facilitating cell division~\cite{Wakefield_2018,Lacroix2018}.
To fulfill these key functions, microtubules must have the ability to rapidly remodel, for both 
assembly and disassembly~\cite{Wakefield_2018}. They achieve this through dynamic instability; periods of
steady growth before switching stochastically into rapid shrinking~\cite{Mitchison1984,Flyvbjerg_1994}. 
The rates of growth, shrinkage, and switching are actively regulated to achieve different 
behaviors~\cite{Lacroix2014,Wakefield_2018,Lacroix2018}.
While it is known that microtubule growth requires GTP hydrolysis~\cite{Flyvbjerg_1994}, and is therefore out of
equilibrium, several competing theoretical models of microtubule dynamics predict different entropy 
production rates~\cite{Howard_2013}. 
\par
By applying our estimators directly to experimental \emph{in vitro} microtubule images,
we can bound the entropy production rate without assuming any particular model for microtubule dynamics.
From a stabilized nucleation site, at constant concentration of tubulin and GTP, microtubules will form, grow, shrink and disappear 
before reforming some time later (Materials and Methods). Kymographs from recent experimental observations~\cite{Lacroix2014} show the steady state trajectories of the microtubule length (Fig.~\ref{Fig:MT}A). Since the length oscillates around its mean value along one spatial 
dimension, the next flux vanishes, so that we have to use $\sigma_2$  to obtain a non-trivial entropy production bound.  Similar to before,
we divide space into 3 periodically layered regions (Fig.~\ref{Fig:MT}B), which yields coarse-grained discretized trajectories as shown in Fig.~\ref{Fig:MT}C. 
Applying $\sigma_2$, we find that a growing microtubule produces entropy at a rate of at least $5 k_B \; min^{-1}$. Furthermore, as demonstrated in Fig.~\ref{Fig:MT}D,E,  choosing a larger spatial or temporal coarse-graining scale decreases the estimates for the entropy production rate. Conversely, this means that higher-resolution experiments promise improved bounds.

\textbf{Induced calcium oscillations in human embryonic kidney cells.}
A coherent cell response to external stimuli requires intracellular signaling~\cite{Thurley2014}. 
One way in which cells encode and transport signal is by controlling the concentration of calcium ions within the 
cytosol~\cite{Sneyd2017,Clapham2007}.  Such calcium oscillations propagate instructions 
for muscle contraction~\cite{Perez2005}, gene expression~\cite{Dolmetsch1998}, and cell differentiation~\cite{Gu1995}.
These oscillations appear as calcium concentration spikes, with $Ca^{+2}$ ions being released into the
cell before ion pumps remove them again~\cite{Sneyd2017,Clapham2007}. Since ion pumps move $Ca^{+2}$ from
a region where the concentration is low (cytosol) to where the concentration is high (sarcoplasmic 
reticulum)~\cite{Sneyd2017}, the system operates out of equilibrium. By measuring the ratio of fluorescence 
at different wavelengths, it is possible to infer the concentration of $Ca^{2+}$ non-invasively within a 
single living cell~\cite{Sneyd2017}. 
\par
In human embryonic kidney cells, calcium oscillations can be triggered by exposure
to carbachol, with the specific response dependent on the concentration of carbachol~\cite{Sneyd2017}. Recent
experiments by Thurley \emph{et al.}~\cite{Thurley2014}, took human embryonic cells and exposed them to a $30\mu M$ 
concentration of carbachol, which after an initial transient resulted in a statistically steady state of 
oscillations (Fig.~\ref{Fig:HEKOsc}A). After an hour, the cells were re-stimulated with a higher concentration of carbachol,
resulting in a new steady state (Fig.~\ref{Fig:HEKOsc}A). As before, we coarse-grain by discretizing the calcium trajectory into three regions, one containing
the default level, one containing intermediate values, and one containing the peaks of the oscillations (Fig.~\ref{Fig:HEKOsc}A).
The coarse-grained trajectories are shown in Fig.~\ref{Fig:HEKOsc}B.  Applying our $\sigma_2$ estimator, we find that prior to stimulation 
the rate of entropy production as at least $4k_B \, min^{-1}$. After exposing the cells to $200\mu M$ carbachol, this bound increases to around $8 k_B \, min^{-1}$
~(Fig.~\ref{Fig:HEKOsc}C).  As in the microtubule case, a finer coarse-graining can be expected to give improved estimates
but will also require a finer temporal resolution than currently available.

\section{Discussion}

\textbf{Entropy production without relative entropies.} It is often implicitly assumed that the best possible bound on entropy production rate comes
from estimating the relative entropies between forward and reverse trajectories, either directly or through TUR~\cite{Horowitz_nocurrent,Hasegawa_2020}. 
To see that non-trivial bounds can be placed on the entropy production rate, even when observable macrostate trajectories appear time
symmetric and so relative entropies are zero, consider a simple Markov chain model of a Brownian clock on four microstates $\{1,2,3,4\}$, with clockwise transition probabilities $q_+$, counter-clockwise
probabilities $q_{-} = 1-q_+$~\cite{Seifert_2016}. When $q_+ > q_-$, the full system is not time-symmetric; for suitably
long observations, a net clockwise current is observed. However, if states 2 and 4 were part of some macrostate $H$, we are just
as likely to observe any forward trajectory on the macrostate set $\{1,3,H\}$ as its time reverse counterpart. To see this, consider an arbitrary observed trajectory,
say $X=(1,H,3,H,3,H)$ of length 6. The probability of observing this macrostate trajectory is $\mathbb{P}(X) = \sum_Y \mathbb{P}(Y)$
where the sum is taken over all microstate trajectories $Y$ consistent with observed macrostate trajectory, which includes $Y = (1,2,3,4,3,4)$. Define $\bar{Y}$ to be the
trajectory where we take $Y$ and switch states 2 and 4, so $\bar{Y} = (1,4,3,2,3,2)$ which has the same macrostate observables. If the microscopic trajectory $Y$ has
$k$ clockwise transitions, and $n$ counter-clockwise transitions,  then $\mathbb{P}(Y) = (1/4) q_+^k q_-^{n-k}$,
whereas $\mathbb{P}(\bar{Y}) = (1/4) q_+^{n-k} q_{-}^{k}$. The time reversed microstate trajectory $Y_r$, has $n-k$ clockwise
transitions and $k$ counter-clockwise, and in general has a different probability of occuring as the forward trajectory. However,
$\mathbb{P}(Y_r) + \mathbb{P}(\bar{Y}_r) = \mathbb{P}(Y) + \mathbb{P}(\bar{Y})$, and so the forward and backward macrostate trajectories are equally probable, $\mathbb{P}(X) = \mathbb{P}(X_r)$. Intuitively,
from the observed statistics, we know that no reversible Markov chain can behave that way; when entering $H$ from 1, trajectories enter
a set of states that typically transition to 3, and vice-versa. Therefore there must be some internal cycles occurring -- even though the
relative entropy of the macroscopic forward and backward trajectories is zero. For the continuous-time version of this example, we can derive analytically the $\sigma_2$-estimator bound, which coincides in this case with  the exact entropy production rate~(SI~\cite{SM}).

\par
\textbf{Range of applicability.}
The optimization framework introduced here can applied to any steady-state meso-scale system that can be modeled by a stationary Markovian (or Langevin-type) dynamics, 
for which the rate of entropy production is related to the relative probability of forward and reverse 
trajectories~\cite{Seifert_2005,Seifert_2012,Horowitz_nocurrent}. These minimal assumptions are fulfilled by many living and active
systems, from single molecules and biomolecular networks~\cite{Seifert_2012}, to molecular motors~\cite{Horowitz_nocurrent},
and active sensors~\cite{Nadrowski2004}. A practical advantage of our method lies in the fact that the coarse-graining level can be adapted to the quality and volume 
of the available experimental data. Here, we focused coarse-graining to a small network with only 3 remaining states, which make it easier to collect precise 
statistics for the transition rates. In general, with  increasing data resolution and trajectory length, finer coarse-graining of space and time will lead to better bounds.
Extrapolating the impressive progress of imaging techniques over the last decade, one can expect that $\sigma_2$-based estimation applied to
higher-resolution data will  enable rapidly improving entropy production rate estimates in the near future.
\par

\textbf{Entropic trade-offs.}
Entropic costs limit the accuracy of biological sensory systems~\cite{Yuhai_2015,Lan_2016}, biological clocks~\cite{Cao2015}, 
and intrinsic noise suppression in cells~\cite{QIAN2006,Paulsson2009Nature}. Beyond direct applications to experimental data, 
the current framework can help us understand and quantify trade-offs between the faithful execution of a biological function 
and the energy expended to do so~\cite{Seifert_2016,Cao2015,Gnesotto_2018}. In particular, since our approach can establish non-trivial 
bounds for a single variable with no observable net currents, it may be used to bound the entropic cost of executing a specific function, such 
as performing oscillations at some frequency and regularity. Furthermore, recent work~\cite{PaulssonPRL2019,Paulsson2009Nature} revealed 
fundamental limits for suppressing molecular fluctuations within cells through negative feedback loops, finding a trade-off 
between control and molecule numbers without  making specific assumptions on the nature of the feedback  loops. 
Similarly, the model-agnostic estimators introduced here could be used to infer additional thermodynamic costs of regulating molecular 
fluctuations by quantifying the entropic trade-offs cells are forced to make.

\section{Conclusion}

Living systems resist their decay into thermal equilibrium by expending entropy to maintain essential cellular processes  and functions~\cite{Paulsson2009Nature,Paulsson2004}. A   quantitative understanding of the associated thermodynamic costs hinges on our ability to infer entropy production rates from partial experimental observations~\cite{Horowitz_nocurrent}.
By recasting this inference problem within an optimization framework, we have constructed an improved rate estimator that can be directly applied to coarse-grained observations of steady-state non-equilibrium systems.  Our analysis of recent experimental data shows that this approach places more accurate bounds on the heat dissipation rates without making specific modeling assumptions. By leveraging information contained in successive transitions,  the derived two-step estimator overcomes a key limitation of previous estimation schemes that require statistically distinguishable forward and reversed trajectories. As a result, we were able to obtain improved bounds on the entropy production of bacterial motors~\cite{Nirody_2019}, microtubules~\cite{Lacroix2014} and calcium oscillations~\cite{Thurley2014}. These successful applications provide guidance for how model-agnostic inference can be used to extract fundamental information from single-variable observations of otherwise hidden intracellular and intercellular processes.

\section{Materials and methods}
\small{
\textbf{Bacteria flagella motor.}
Bacterial flagella bead trajectories were provided Jasmine Nirody, and obtained similarly to Nirody \emph{et al.}~\cite{Nirody_2019}.
In their recent experiments, \emph{E. coli} bacteria were immobilized on a cover slip, and a bead ($1 \mu$m) was attached to their
shortened flagella. The bead position is found using back focal-plane interferometry~\cite{Nirody_2019,BerryScience,KRASNOPEEVA2019}. 
The strains MTB24 and MTB32 were used, with the motor powered by sodium ions in both cases. Concentrations  were $10mM$ and $85mM$ $Na^{+}$ for MTB24, and  $85mM$ $Na^{+}$  for MTB32. 
A single trajectory of length 20s was taken from each experiment, with each 20s window containing at least 400 rotations. 
For each trajectory, the origin was taken as the trajectory center of mass in the $xy$ plane. The plane was then divided into 3 regions made 
from $3N$ segments, with $N$ chosen for each trajectory to maximize the entropy production rate bound. In total, we  analyzed 7 MTB24 $10mM$ 
trajectories, 25 MTB24 $85mM$ trajectories, and 10 MTB32 $85mM$ trajectories.\\
}

\small{
\textbf{Microtubule dynamic instability.}
Experimentally measured microtubule trajectories were provided by Benjamin Lacroix. 
Stabilized guanylyl 5$'$-$\alpha$,$\beta$-methylenediphosphonate (GMPCPP) seeds were attached to a functionalized surface and served as nucleation sites.
They were placed in a solution of $7\mu M$ tubulin and  $1mM$ guanosine triphosphate (GTP) at a temperature of  $35^{\circ} C$.
The growing microtubules were imaged by total internal reflection fluorescence (TIRF) microscopy, and kymographs were automatically extracted, 
from which the microtubule length was calculated. Data from 2 experiments performed under identical conditions were used in our analysis, with
1200 minutes of total observation time.\\
}

\small{
\textbf{Calcium oscillations.}
The calcium concentration trajectories were taken from recent experiments by Thurley \emph{et al.}~\cite{Thurley2014}, with 20
trajectories for the protocol of $30\mu M$ carbachol stimulation followed by $100\mu M$ carbachol re-stimulation, and 14 trajectories for 
$30\mu M$ carbachol stimulation followed by $200 \mu M$ carbachol re-stimulation. The same concentration coarse-graining into 3 regions 
was applied to data takenat $30 \mu M$ and $100\mu M$ carbachol. The coarse grained states were adapted for data corresponding the $200\mu M$ carbachol 
re-stimulation, as these tended to be larger in amplitude and displayed higher fluorescence ratio between spikes.
}

\section{Acknowledgements}
We thank Jasmine Nirody for providing the bacterial flagella trajectories, Benjamin Lacroix for sharing the microtubule trajectories and
Alexander Skupin for providing the calcium oscillation data, and all of them for explaining their experiments to us.
We are also grateful to  Massimiliano Esposito and Jordan Horowitz for helpful discussions and insightful comments on an early 
manuscript draft. This work was supported by a MathWorks Fellowship (D.J.S.), a James S. McDonnell Foundation Complex 
Systems Scholar Award (J.D.), and the Robert E. Collins Distinguished Scholar Fund (J.D.).
\clearpage

\onecolumngrid
\hypersetup{
  colorlinks   = true, 
  urlcolor     = blue, 
  linkcolor    = black, 
  citecolor   = black 
}

\renewcommand{\d}{\text{d}}
\renewcommand{\div}[2]{\frac{\d {#1}}{\d {#2}}}
\renewcommand{\v}{\boldsymbol}
\setcounter{equation}{0}
\setcounter{figure}{0}
\setcounter{table}{0}
\setcounter{page}{1}
\renewcommand{\theequation}{A\arabic{equation}}
\renewcommand{\thefigure}{A\arabic{figure}}
\renewcommand{\bibnumfmt}[1]{[A#1]}
\renewcommand{\citenumfont}[1]{A#1}

\begin{center}
  \textbf{\large Appendix: Improved bounds on entropy production in living systems}\\[.2cm]
Dominic J. Skinner,$^{1}$ and J\"{o}rn Dunkel$^1$\\[.1cm]
 {\itshape ${}^1$Department of Mathematics, Massachusetts Institute of Technology, Cambridge, Massachusetts 02139-4307, USA\\}
(Dated: \today)\\[1cm]
\end{center}

\section{Notation and preliminaries}
For a detailed introduction to continuous time Markov processes on a discrete set of states, we refer to Ref.~\cite{Svan1992stochastic}.
Here, we consider a continuous time Markov process, $X_t$, on $N$ states with generator $Q = (q_{ij})$, so that
\begin{equation}
\mathbb{P}(X_t = j | X_0 = i) = P_{ij}(t) = (e^{Qt})_{ij},
\end{equation}
where $\sum_j q_{ij} = 0$, $q_{ij}$ represents the transition rate from state $i$ to state $j$, and 
$-q_{ii} = \sum_{j\neq i}q_{ij}$ the rate at which trajectories leave state $i$. Given a distribution 
$\mu(t)=(\mu_1(t),\ldots,\mu_n(t))$, its time evolution satisfies
\begin{equation}
\dot{\mu}_i(t) = \sum_{j\neq i} \mu_j q_{ij} - \sum_{j\neq i} \mu_i q_{ij} = \sum_j \mu_j q_{ji},
\end{equation}
and hence finding a stationary distribution corresponds to finding a left zero--eigenvector,
$\pi Q = 0$. 
For a given system, we want to quantify the steady-state rate of entropy production~\cite{SGingrich_natPhys}, 
\begin{equation}\label{eq:EntMC}
\sigma = \frac{k_B}{2} \sum_{i \neq j} (\pi_i q_{ij} - \pi_j q_{ji} ) \log \left( \frac{\pi_i q_{ij}}{\pi_j q_{ji}}\right),
\end{equation}
where we take the sum over $(i,j)$ where $q_{ij} > 0$. The rate of entropy production can also be written in terms of
the relative ratio of forward and reverse trajectories,
\be\label{eq:Ent}
\sigma = \lim_{T \to \infty} \frac{k_B}{T} \left\langle \log 
\frac{ \mathbb{P}_f(X_t)}{\mathbb{P}_r(X_t)} \right\rangle,
\ee
where $\mathbb{P}_f(X_t)$ is the probability of observing the forward path $X_t$, $\mathbb{P}_r(X_t)$ is the
probability of observing the reverse path $Y_t = X_{T-t}$, and the expectation is taken over all 
paths of length $T$~\cite{SHorowitz_nocurrent}. For Markovian systems in steady state, equations~\eqref{eq:EntMC}
and~\eqref{eq:Ent} are equivalent. We are also assuming the underlying Hamiltonian is an even function with
respect to inversion of momenta~\cite{SParrondo_2009}.

The first definition of the entropy production rate in Eq.~\eqref{eq:EntMC} suggests an alternative formulation of the system 
by defining  $n_{ij} = \pi_i q_{ij}$
for $i\neq j$,  which must satisfy conservation of mass at each $j$,
\begin{align}
\sum_{i \neq j} n_{ij} &= \sum_{i \neq j} n_{ji},
\end{align}
together with the normalization condition
\begin{equation}
\sum_i \pi_i = 1
\end{equation}
 for the stationary distribution $\pi_i\ge 0$. The rate of entropy production becomes
\begin{equation}
\sigma = \frac{k_B}{2} \sum_{i\neq j} (n_{ij} - n_{ji} ) \log \left( \frac{n_{ij}}{n_{ji}}\right).
\end{equation}
We will use this formulation from now on, and typically will not explicity specify $\pi_i$.

For Markovian systems, upon arriving at $i$, an exponential waiting time begins with probability 
density of jumping at $t$ being,
\begin{equation}
\psi_i(t)\mathrm{d}t := \mathbb{P}(X_{t+\mathrm{d}t} \neq i | X_{t} = i)=
 \lambda_i e^{-\lambda_i t}\mathrm{d}t, 
\end{equation}
with $\lambda_i = \sum_{j\neq i} n_{ij} /\pi_i$. 

We take a moment to note that the function,
\begin{equation}
f(x,y) = \logfn{x}{y},
\end{equation}
satisfies $f(x,y) = f(y,x) \geq 0$ for $x, y > 0$, with $f(x,y) = 0$ only when $x=y$. Further, by examining 
the Hessian,
\begin{equation}
Hf = \left(\frac{1}{x} + \frac{1}{y} \right) \left[ \begin{array}{cc}
y/x & -1 \\ -1 & x/y \end{array} \right],
\end{equation}
we see $Det(Hf) =0$, $Tr(Hf) > 0$, making $Hf$ positive semi-definite, and hence meaning $f$ is convex on 
the region $x,y >0$.

\section{Observed system is not $N^{th}$ order Markovian}
\begin{figure}\centering
\includegraphics{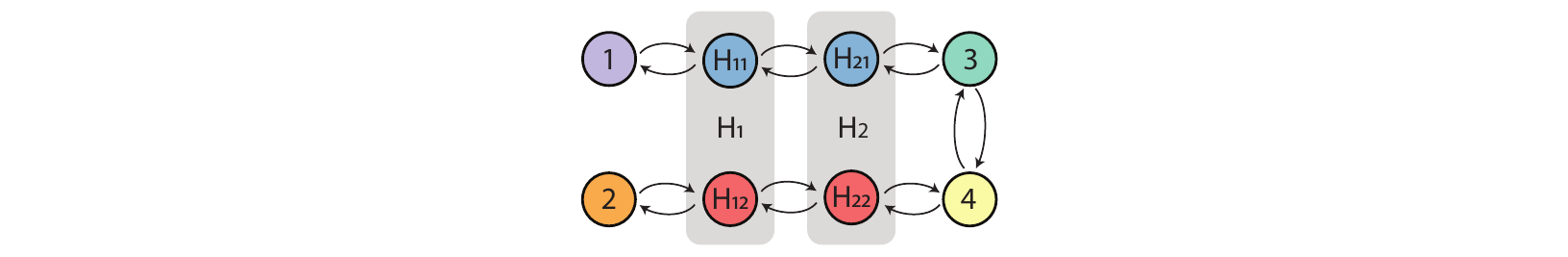}
\caption{\label{fig:NonMarkov} Example system demonstrating that observed macrostates are not $N^{th}$ order Markovian.}
\end{figure}
While the underlying system is Markovian, the observed system is not. For instance, consider the system
in Fig.~\ref{fig:NonMarkov}. Suppose the system is currently in macrostate $H_1$. Without any
other knowledge, the probabilities of transitioning to states 1 or 2 are both non-zero. However, 
if the previous observed macrostate was $1$, the probability of transitioning to 2 in the next
jump is exactly zero, showing that past information is important and hence the observed system
is not Markovian. Further, given the observed trajectory of length $N+1$,
$(3, H_{2}, H_{1},H_2, \dots, H_1)$, we know the probability of transitioning to state $2$ is zero in the next 
jump, but if we are only given the shortened trajectory
$(H_2, H_{1},H_2, \dots, H_1)$, we do not know this, showing that trajectories cannot be 
truncated to length $N$ without losing information, and hence this system is not $N^{th}$ order Markovian.

\section{The estimator $\sigma_1$}
\begin{figure}\centering
\includegraphics{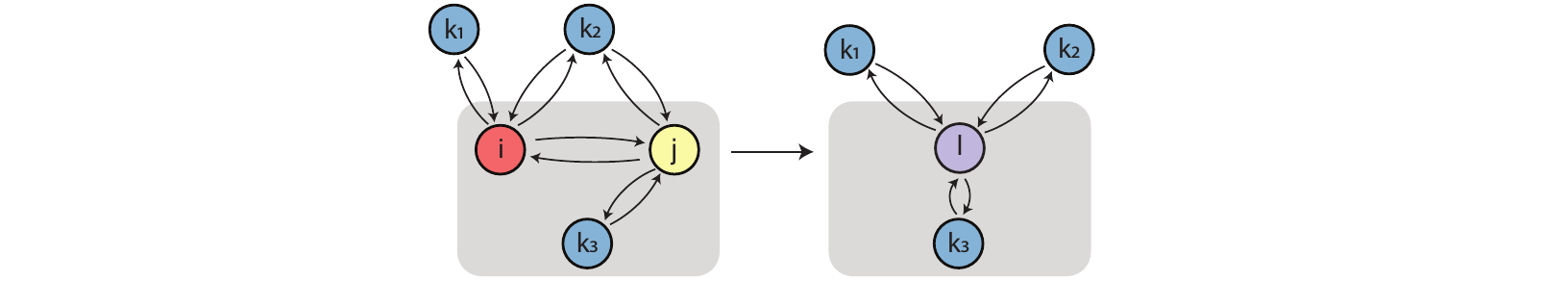}
\caption{\label{fig:O1}Simplifying the internal topology without affecting the observables $\mathcal{O}_1 = \{ n_{IJ} \}$.
Joining the states $i$, $j$ within the same macrostate to form a new state $l$ lowers the entropy whilst
not altering $\mathcal{O}_1$-statistics. }
\end{figure}
To construct the estimator $\sigma_1$ we show that given any system $\mathcal{R}$ consistent with 
$\mathcal{O}_1=\{n_{IJ} | I,J \text{ macrostates}\}$,
we can combine two hidden states in the same macrostate
in such a way that  the rate of entropy production is lowered whilst not affecting the observables in $\mathcal{O}_1$.
Specifically, consider two states $i$, $j$ within the same macrostate, with local arrival rates $n_{ki}$, $n_{kj}$,
and departure rates $n_{ik}$, $n_{jk}$, Fig.~\ref{fig:O1}. Now replace $i$, $j$ by a single state $l$ with arrival rates
$n_{kl} = n_{ki} + n_{kj}$, departure rates $n_{lk} = n_{ik} + n_{jk}$, and mass density 
$\pi_l = \pi_i + \pi_j$, Fig.~\ref{fig:O1}. This does not alter any of the observables
in $\mathcal{O}_1$, since it preserves the mass flux between macrostates. Consider now the effect on
the rate of entropy production. The entropy produced along the edge connecting states $i$ and $j$ vanishes
after this change. The entropy production rate along edges connecting $k$ to $i$ and $j$ only changes if both $n_{ik}$
and $n_{jk}$ were non zero, in which case it changes as
\begin{equation}
(n_{ik} - n_{ki}) \log \left( \frac{n_{ik}}{n_{ki}} \right) + 
(n_{jk} - n_{kj}) \log \left( \frac{n_{jk}}{n_{kj}} \right)
\quad \mapsto \quad
((n_{ik} + n_{jk}) - (n_{ki} + n_{kj})) \log \left( \frac{n_{ik}+n_{jk}}{n_{ki}+n_{kj}} \right).
\end{equation}
This reduces the entropy because $f(x,y) = (x-y)\log(x/y)$ is a convex function. 
Therefore, combining $i$ and $j$ into the new state $l$ has not increased the total
rate of entropy production. Thus given any $\mathcal{R}$ that satisfies $\mathcal{O}_1(\mathcal{R}) = \mathcal{O}_1
(\mathcal{S})$, we can apply this procedure iteratively, which will not increase the entropy production rate,
until each macrostate only contains one hidden state. 
The resulting system is  a Markovian system with no hidden states and the same transition rates
as the observed transition rates between macrostates. The entropy production rate of this system is the $\sigma_1$
estimator.

\section{The estimator $\sigma_2$}
\begin{figure}
\includegraphics{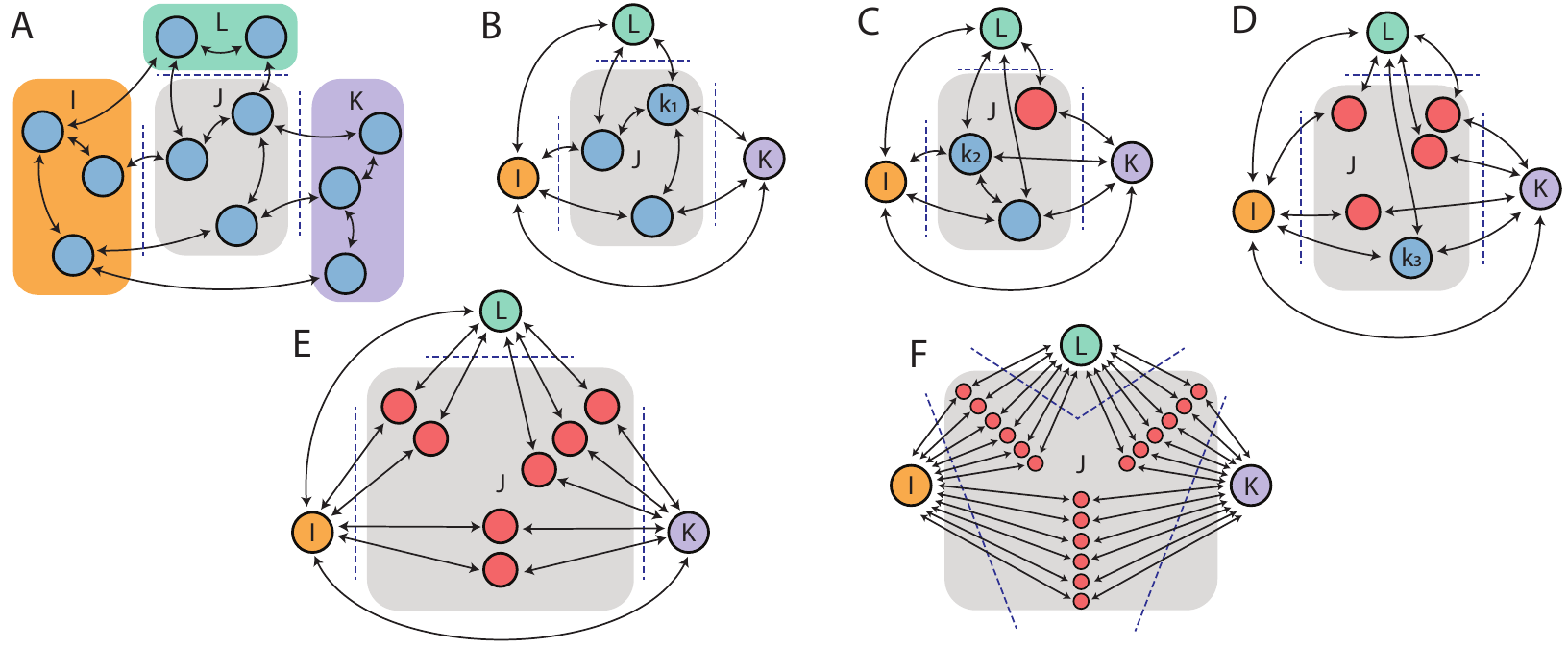}
\caption{\label{fig:O2}Transforming an arbitrary network topology into a canonical form, whilst lowering the
entropy production rate, and preserving statistics involving $J$. (A) An arbitrary network, for which we are 
finding the minimal entropy production rate across the edges $(J,I)$, $(J,K)$, and $(J,L)$, indicated by dashed lines,
subject to the statistics
$n_{IJ}$, $n_{JI}$, $n_{JK}$, $n_{KJ}$, $n_{JL}$, $n_{LJ}$, $n_{IJK}$, $n_{KJI}$, $n_{IJL}$, $n_{LJI}$,
$n_{KJL}$, and $n_{LJK}$. (B) We apply the iterative procedure as for the
$\sigma_1$ estimator, to combine all the microstates in $I$ into a single state, similar for $K$, $L$. 
(C) The rerouting procedure is applied to the microstate previously labeled $j_1$ resulting in a new microstate (red)
that has no internal connections within $J$. 
(D) The rerouting procedure is applied to the microstate previously labeled $j_2$, resulting in 3 new 
microstates (red) that have no internal connections within $J$.
(E) The rerouting procedure is applied to the microstate previously labeled $j_3$, resulting in
3 new microstates (red), so now
there are no internal connections within~$J$. This is the canonical form, but while in the shown example there are 2 microstates
connecting $I$ and $K$, in theory there could be arbitrarily many. (F) We prove later that when performing
the numerical minimization, only 6 states per pair of external macrostates are needed, as shown here.}
\end{figure}

To construct the estimator $\sigma_2$, we first bound the rate of entropy production along all edges connected
to a macrostate $J$, whilst preserving the local observed quantities
$\{n_{IJ}, n_{IJK}\}$, for all distinct macrostates $I$, $K$ that are neighbors of $J$, a subset of the global
set of $\mathcal{O}_2$-observables, $\mathcal{O}_2 = \{ n_{UV}, n_{UVW} | U,V,W \text{ macrostates} \}$.
To obtain such a local bound for each macrostate $J$, we take any system consistent with the local $\mathcal{O}_2$-statistics, and show
it can be transformed into a canonical form. We then find the local bound by optimizing numerically 
over this canonical form. Finally, we will combine the local bounds of all macrostates to construct the 
global bound $\sigma_2$.
We outline the steps here, with a worked example shown in Fig.~\ref{fig:O2}.

\textbf{Simplify topology for states other than $J$:} 
First, we combine microstates within all neighboring macrostates of $J$, by the procedure outlined for the 
$\sigma_1$ estimator. This may decrease the entropy production rate on the edges we are concerned with, 
but importantly it will not change the local $\mathcal{O}_2$-statistics which we aim to preserve. To see this, we first note
that the procedure preserves $\mathcal{O}_1$ statistics by construction. Second, it does not affect quantities
like $n_{IJK}$, since paths still arrive and leave the microstates of $J$ at the same rate as before, and the internal
topology of $J$, which is preserved, determines the conditional probability of where those paths will end up.
This simplification step corresponds to the transformation of Fig.~\ref{fig:O2}A to Fig.~\ref{fig:O2}B.

\textbf{Rerouting procedure for microstates in $J$:} 
We now consider a state $j$ within the macrostate $J$ and suppose it is connected to various states $i$, $k$
which may or may not be within $J$. Let the neighboring states of $j$ be the set $\mathcal{K}$, so the entropy
production along edges connected to $j$ is 
\begin{equation}
\sigma(j) = \sum_{i \in \mathcal{K}} \logfn{n_{ij}}{n_{ji}}.
\end{equation}
For states $i$ and $k$, the rate at which mass leaves $i$ to go to $j$ and then subsequently to $k$
is $\tilde{n}_{ik} = n_{ij}n_{jk}/\mathcal{N}$, where $\mathcal{N} = \sum_l n_{jl}$. Instead of sending
this mass through the path $i\to j\to k$ we could instead send it directly from $i\to k$. We will see
that this lowers the entropy production rate, and if at least one of $i$, $k$ is in $J$ it preserves
both the rate at which mass flows in/out of $J$, as well as the conditional statistics, since only
the intermediate state of a path is removed. If neither
$i$ nor $k$ are in $J$, sending paths straight from $i\to k$ would change the macrostate path statistics,
which we intended to preserve. In this case, we send mass through some new intermediate state $l$ within $J$.
The effect of this rerouting is shown in Fig.~\ref{fig:O2}B-C.
To preserve the $\mathcal{O}_2$-statistics, the new rate of mass transfer should maintain the rate at which 
mass ends up back at $i$ or at $k$, so that
\begin{equation}\arraycolsep=6.4pt
\begin{array}{cc}
\hat{n}_{il} = n_{ij}(n_{ji} + n_{jk}) /\mathcal{N}, &\qquad \hat{n}_{jl} = n_{kj}(n_{ji}+n_{jk})/\mathcal{N}, \\[4pt]
\hat{n}_{li} = n_{ji}(n_{ij} + n_{kj}) /\mathcal{N}, &\qquad \hat{n}_{lk} = n_{jk}(n_{ij}+n_{kj})/\mathcal{N}.
\end{array}
\end{equation}
Consider now the effect on the rate of entropy production. Since
\begin{equation}
\logfn{\hat{n}_{il}}{\hat{n}_{li}} + \logfn{\hat{n}_{kl}}{\hat{n}_{lk}} = 
\frac{1}{\mathcal{N}}\logfn{n_{ij}n_{jk}}{n_{kj}n_{ji}},
\end{equation}
the inclusion of the intermediate state ends up producing the same rate of entropy as if the mass
was directly rerouted, with $\tilde{n}_{ik}$.
Therefore, the total rate of entropy production on the new edges will be,
\begin{align}
&\frac{1}{2}\sum_{i,k \in \mathcal{K}} 
\frac{1}{\mathcal{N}}\logfn{n_{ij}n_{jk}}{n_{kj}n_{ji}} 
\nonumber\\
\nonumber
&= \frac{1}{2} \sum_{i \in \mathcal{K}} 
\frac{1}{\mathcal{N}}\left( n_{ij} \sum_{k\in \mathcal{K}} n_{jk} - n_{ji} \sum_{k\in \mathcal{K}}n_{kj} \right) 
\log \left( \frac{n_{ij}}{n_{ji}} \right) + 
\sum_{k \in \mathcal{K}} 
\frac{1}{\mathcal{N}}\left( n_{kj} \sum_{i\in \mathcal{K}} n_{ji} - n_{jk} \sum_{i\in \mathcal{K}}n_{ij} \right) 
\log \left( \frac{n_{kj}}{n_{jk}} \right) 
\\\nonumber
&= \frac{1}{2}\sum_{i\in\mathcal{K}}\logfn{n_{ij}}{n_{ji}} + \frac{1}{2}\sum_{i\in\mathcal{K}}
\logfn{n_{kj}}{n_{jk}} 
\\
&= \sum_{i\in\mathcal{K}}\logfn{n_{ij}}{n_{ji}} 
\notag\\
&= \sigma(j).
\end{align}
thus, the same as the original entropy production rate $\sigma(j)$.
However, if there was already mass transport directly from some
$i\to k$ which now has additional mass transport after the rerouting, the entropy production rate changes as we
combine these two mass transport paths. In this case, due to convexity in
the entropy production rate function, the entropy production rate will not increase after the rerouting.

\textbf{Iterative rerouting:}
By repeatedly applying the rerouting procedure, all internal edges can be removed from $J$, and
every microstate connects to exactly 2 external macrostate, see Fig.~\ref{fig:O2}C-F. 
This could still leave an arbitrary number of microstates in $J$ connecting, for instance, $I$ and $K$. In the next section, we prove that minimizing over 6 microstates sufficies, 
so that we can numerically perform the minimization to find the bound.

\textbf{Proving that 6 states suffices:}
Suppose we have minimized over all internal topologies with a maximum of $\tilde{N}$ internal states,
consistent with the local $\mathcal{O}_2$-statistics involving $J$, and transformed this into the canonical form. 
There will be $N\leq \tilde{N}$ internal states of $J$ that connect $I$ and $K$. Since the resulting canonical system already has the minimal entropy production rate,
we cannot tweak the transition rates, consistent with the constraints, to lower the entropy production rate further.
This fact will allow us to prove that we only need at most $N=6$ internal states in $J$ that connect $I$ and $K$ to get the same bound. 
\par
To show this, we label the relevant transition rates $n_{Ij}$, $n_{jI}$, $n_{Kj}$, $n_{jK}$, where $j$ indexes the $N$ internal
states in $J$ that connect $I$ and $K$. To preserve the macrostate transition rates, we must preserve
\begin{subequations}
\label{e:constraints}
\begin{equation}
\sum_{j=1}^N n_{Ij} = C_1, \qquad \sum_{j=1}^N n_{jI} = C_2, \qquad \sum_{j=1}^N n_{Kj} = C_3, \qquad \sum_{j=1}^N
n_{jK} = C_4,
\end{equation}
where $C_1\neq n_{IJ}$ in general, since we are only considering the microstates connecting $I$ to $K$, and not
other microstates. To preserve the conditional transition rates, we have that
\begin{equation}
\sum_{j=1}^N \frac{n_{Ij}n_{jK}}{n_{jI} + n_{jK}} = n_{IJK}, \quad\qquad
\sum_{j=1}^N \frac{n_{Kj}n_{jI}}{n_{jI} + n_{jK}} = n_{KJI}.
\end{equation}
\end{subequations}
Conservation of mass also requires 
$n_{jI} + n_{jK} = n_{Ij} + n_{Kj}$. The rate of entropy production on these edges is given by
\begin{equation}
\sigma_{IJK} = \sum_j \left[ \logfn{n_{jK}}{n_{Kj}} + \logfn{n_{jI}}{n_{Ij}}\right] ,
\end{equation}
which has been minimized under these constraints. \\

\par
To reduce the number of internal states $N$, first note that if for some state $n_{jK} = 0$, then thermodynamic reversibility~\cite{SSeifert_AnnRev}  requires $n_{Kj} =0$ as well.
If additionally $n_{jI}=0$, then the state can be removed so assume in this case
$n_{jI}, n_{Ij} \neq 0$. 
If there are multiple of these states, simply combine them, which will not affect the linear transition
rate statistics, nor the conditional statistics which will have a zero contribution from these edges
anyway. Hence we need at most one of these states for $n_{jK}=0$, and one for $n_{jI}=0$, and 
so the remaining states have all of $n_{jI}$, $n_{Ij}$, $n_{jK}$, $n_{Kj}$ non-zero, labeling 
these states as $1,\dots, M$, where $N-2 \leq M$. \\

We now optimize over the remaining states, defining the mass arriving
at each internal state as $\lambda_j$, so that 
\be
\lambda_j = n_{Ij} + n_{Kj} = n_{jI} + n_{jK},\quad \text{ for } j = 1,\dots,M,
\ee
and let $\lambda_j y_j = n_{jK}$, $\lambda_j z_j = n_{Kj}$, from which it follows that
$n_{jI} = \lambda_j (1-y_j)$, $n_{Ij} = \lambda_j (1-z_j)$. At our minima, we know that
$\lambda_j >0$, and $x_j,y_j \in (0,1)$. With $x = (\lambda_1,\dots,\lambda_M, y_{1},\dots, y_{M}, 
z_{1}, \dots, z_{M})$, we have minimized
\be
f(x) =\sigma_{IJK} =  \sum_{j}\lambda_j (y_{j} - z_{j}) \log \left( \frac{y_{j}(1-z_{j})}{z_{j}(1-y_{j})} \right).
\ee
Conservation of mass is automatically enforced in these
variables, and the constraints Eq.~\eqref{e:constraints} can be
 written as $c_i(x) = 0$, with
 \begin{subequations}
\begin{align}
c_1(x) &= \sum_j \lambda_j y_{j} - \hat{c}_1 = 0\\
c_2(x) &= \sum_j \lambda_j z_{j} - \hat{c}_2 = 0\\
c_3(x) &= \sum_j \lambda_j - \hat{c}_3 = 0\\
c_4(x) &= \sum_j \lambda_j y_{j} z_{j} - \hat{c}_4 = 0,
\end{align}
\end{subequations}
for some constants $\hat{c}$. We calculate the gradient of $f$, $g_k(x) = \partial_k f$, to obtain
 \begin{subequations}
\begin{align}
g_j(x) &= (y_{j} - z_{j}) \log \left( \frac{y_{j}(1-z_{j})}{z_{j}(1-y_{j})} \right), \\
g_{j+M}(x) &= \lambda_j \left[ \log \left( \frac{y_{j}(1-z_{j})}{z_{j}(1-y_{j})} \right) +
\frac{y_{j} - z_{j}}{y_{j}(1-y_{j})} \right], \\
g_{j+2M}(x) &= \lambda_j \left[ \log \left( \frac{z_{j}(1-y_{j})}{y_{j}(1-z_{j})} \right) +
\frac{z_{j} - y_{j}}{z_{j}(1-z_{j})} \right], 
\end{align}
\end{subequations}
for $j = 1,\dots M$. The gradient vector of the constraints, $a_i = \nabla c_i$, read
 \begin{subequations}
 \begin{align}
a_1(x) &= \left[y_{1} , \dots, y_{M}, \lambda_1 , \dots, \lambda_M, 0, \dots, 0 \right], \\
a_2(x) &= \left[z_{1}, \dots, z_{M},0, \dots, 0 ,  \lambda_1 , \dots, \lambda_N \right], \\
a_3(x) &= \left[1 , \dots, 1, 0, \dots, 0, 0, \dots, 0 \right], \\
a_4(x) &= \left[y_{1}z_{1} , \dots, y_{M}z_{M}, \lambda_1z_{1} , \dots, \lambda_M z_{M}, 
\lambda_1 y_{1}, \dots, \lambda_M y_{M} \right], 
\end{align}
\end{subequations}
which are the columns of the Jacobian $A=[a_1^\top, a_2^\top, a_3^\top, a_4^\top]$. We note that the only way these columns would
not be full rank is if $y_{i} = const.$, $z_{i} = const.$, at which point we could combine
all $M$ states into a single state without altering the statistics. Supposing that the Jacobian $A$
is full rank, we can make use of the following necessary condition~\cite{Sliberzon2011calculus}:
\begin{theorem*}\nonumber
A necessary conditions for $x$ to be a local minimizer is that 
$g(x) = A(x)^\top \mu$ for some $\mu$.
\end{theorem*}
This condition will limit us to at most 4 choices of $y_j$, $z_j$, showing that we could combine
the $M$ internal states to at most 4 states.
This condition implies that
\begin{align}
(y_j - z_j)\log \left( \frac{y_j (1-z_j)}{z_j (1-y_j)} \right) &= 
\mu_1 y_j + \mu_2 z_j +\mu_3 + \mu_4 y_j z_j, 
\notag\\
\log \left( \frac{y_j (1-z_j)}{z_j (1-y_j)} \right) + \frac{y_j-z_j}{y_j(1-y_j)} &= 
\mu_1 +  \mu_4 z_j, 
\notag\\
-\log \left( \frac{y_j (1-z_j)}{z_j (1-y_j)} \right) - \frac{y_j-z_j}{z_j(1-z_j)} &= 
\mu_2 +  \mu_4 y_j,
\notag
\end{align}
where we have divided by $\lambda_j> 0$. Eliminating the log term gives,
\begin{align}
(y_j - z_j) \left[ \mu_1 + \mu_4 z_j - \frac{y_j - z_j}{y_j(1-y_j)} \right] &=
\mu_1 y_j + \mu_2 z_j +\mu_3 + \mu_4 y_j z_j, 
\notag\\
(y_j-z_j)\left[ \frac{1}{y_j(1-y_j)}- \frac{1}{z_j(1-z_j)} \right]  &= 
\mu_1 + \mu_2 + \mu_4 (y_j + z_j), 
\notag
\end{align}
which in turn imply that 
\begin{align}
(y_j - z_j) \left[ - \frac{y_j - z_j}{y_j(1-y_j)} \right] &=
(\mu_1+\mu_2) z_j + \mu_3 + \mu_4 z_j^2, 
\notag\\
z_j(y_j-z_j)\left[ \frac{1}{y_j(1-y_j)}- \frac{1}{z_j(1-z_j)} \right]  &= 
(\mu_1 + \mu_2)z_j + \mu_4 (y_j + z_j)z_j.
\notag
\end{align}
From these relations, we deduce that
\begin{align}
\frac{(y_j - z_j)^2 }{(1-y_j)(1-z_j)} &= \mu_4 y_j z_j - \mu_3 
\notag\\
\frac{(y_j - z_j)^2 }{(1-y_j)(1-z_j)}(y_j + z_j -1) &= y_jz_j 
\left[\mu_1 + \mu_2 + \mu_4(y_j+z_j)\right],
\notag
\end{align}
which also implies that
\be
(\mu_1 + \mu_2 + \mu_4)y_jz_j + \mu_3(y_j + z_j - 1) = 0.
\ee

\paragraph{Case 1:}
If $\mu_1 + \mu_2 + \mu_4 = 0$, then $1 = y_j + z_j$, so
\be
\frac{(2 - y_j)^2}{y_j(1-y_j)} = \mu_4 y_j (1-y_j) -\mu_3,
\ee
which is a quartic equation in $y_j$, with at most two solutions in $(0,1)$.

\paragraph{Case 2:}
If $\mu_1 + \mu_2 + \mu_4 \neq 0$, then
\be 
y_jz_j = \alpha (1 - y_j -z_j),
\ee
with $\alpha = \mu_3/(\mu_1 + \mu_2 + \mu_4)$, 
which is a hyperbola with a single branch in $(0,1)^2$ for $\alpha < -1$, or $\alpha > 0$.
Substituting into the other constraint yields a quartic,
\begin{align}
-\alpha^2 + (4 \alpha^2 &- \alpha \mu_3 - \alpha^2 \mu_3) y_j + (2 \alpha - 4 \alpha^2 
+ \alpha \mu_4 + \alpha^2 \mu_4 - \mu_3 + \alpha^2 \mu_3) y_j^2 \\ \nonumber
& + (-4 \alpha - 2 \alpha \mu_4 - 2 \alpha^2 \mu_4 + \mu_3 + \alpha \mu_3) y_j^3 + (-1 + \alpha \mu_4 + 
    \alpha^2 \mu_4) y_j^4 = 0,
\end{align}
 which can have at most 4 unique solutions. This means there are at most $4$ solutions for $(y_j,z_j)$ values, 
but if two internal states have the same value of $(y_j,z_j)$, they can be combined without affecting
the $\mathcal{O}_2$ statistics or the entropy production rate.
Therefore, if $M > 4$, we simply combine states until there are at most 4 remaining internal states, all
with unique values of $(y_j,z_j)$. Hence, to find the minimum value,  $M=4$ and so $N=6$ states are sufficient.

\textbf{Constructing full estimator:}
We have outlined how to bound the local entropy production rate $\sigma_2(J)$ for the edges connected to a single
macrostate $J$. Constructing this bound for every observed macrostate $J$, defines the global estimator $\sigma_2$, 
\begin{equation}
\sigma_2 = \frac{1}{2} \sum_{J} \sigma_2 (J) \leq \frac{1}{2} \sum_{J} \sum_{I} \sigma(I,J) \leq \sigma,
\end{equation}
where $\sigma(I,J)$ is the entropy produced over all edges connecting $I$ to $J$, with 
every edge bounded twice. For a simple 3 state topology, on states $A$, $B$, $C$, with no direct 
transitions between macrostates $A$ and $C$, the estimate $\sigma_2(B)$ is the optimal 
bound given all $\mathcal{O}_2$ statistics. We use this estimator for the analytic verification
section, and for the Calcium oscillation example in the main text. Elsewhere, when this topology
cannot be assumed, we use the $\sigma_2$ estimator.

\subsection{Numerical implementation}
The above analytic results guarantee that we can find the minimum entropy producing state, consistent with
the statistics, by numerically optimizing over a known, and finite, network topology. 
To compute the numerical solutions to this problem under the constraints~\eqref{e:constraints}, we used a global search non-linear optimization
algorithm~\cite{SGlobalSearch}. The problem was solved in terms of the variables $n_{jK}$,
only specifying 3 out of the 4 rates for an internal state with the remaining fourth rate then determined by mass conservation.

\section{Analytic verification}
\begin{figure}\centering
\includegraphics{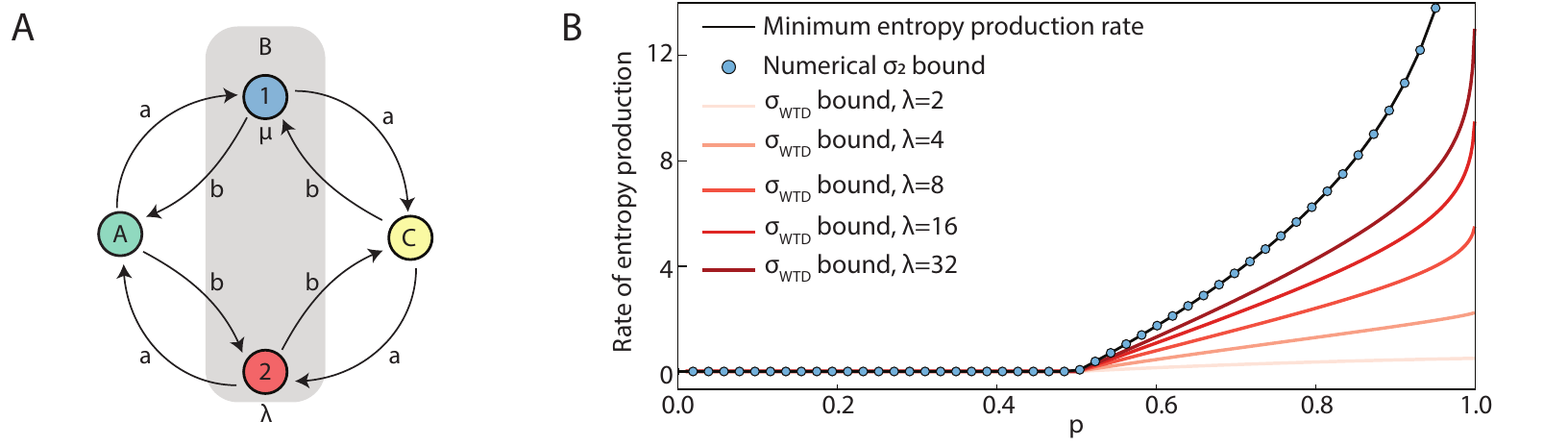}
\caption{\label{fig:AnalyticMin} (A) An example system that minimizes the entropy production rate subject to 
$n_{AB} = n_{BA} = n_{BC} = n_{CB} = 1$, and $n_{ABC} = p \in [0,1)$. The internal states
are assumed to have waiting time parameters $\lambda$ and $\mu$, respectively, although these do not affect the entropy
production. (B) Comparison of the global optimization, the analytic minimum, and a waiting time distribution bound.
The $\sigma_2$-optimization consistently finds the global minimum for these parameters. As the waiting time distributions with $\mu=1$ and
$\lambda>1$ become more distinct, the waiting time estimator improves, but requires an order of magnitude difference 
to get a reasonable bound.}
\end{figure}

In general, analytic solutions of the minimization problem are not readily available. However, for special symmetric
cases a bound can be calculated analytically by using an ansatz respecting the symmetry constraints. 
To demonstrate this, we consider a system
with three states where $n_{AB} = n_{BA} = n_{BC} = n_{CB} = 1$, and $n_{ABC} = p \in [0,1)$,
the control parameter. The symmetry ansatz assumes the transitions $n_{A1} = a$, $n_{1A} = b$, $n_{A2} = b$,
$n_{2A} = a$, $n_{1C} = a$, $n_{C1} = b$, $n_{2C} = b$, $n_{C2}=a$, with 2 internal states, Fig.~\ref{fig:AnalyticMin}A. 
The optimization problem to be solved then reads
\be
\min_{a,b} \; \sigma = 4\logfn{a}{b}
 \qquad \text{subject to \quad} \begin{array}{c} a+b \leq 1 \\[4pt]
 \frac{a^2+b^2}{a+b} = p \end{array}
\ee 
This has a solution producing zero entropy as long as $p \leq \frac{1}{2}$. If $p>\frac{1}{2}$, then
the optimal solution has $a+b = 1$, and so (taking $a>b$), 
\be 
a = \frac{1 + \sqrt{2p-1}}{2}, \qquad
b = \frac{1 - \sqrt{2p-1}}{2}, \qquad
\sigma = 4 \sqrt{2p-1} \log \left( \frac{ 1+ \sqrt{2p-1}}{1 - \sqrt{2p-1}} \right),
\ee

Applying the numerical optimization to the observed statistics, finding a bound on the entropy produced on edges
neighboring the state $B$, we are able to find the minimum, Fig.~\ref{fig:AnalyticMin}B. 

\section{Time symmetric observables and comparison with waiting time estimator}
For the system depicted in Fig.~\ref{fig:AnalyticMin}A, if the waiting time parameters are identical, $\mu = \lambda$, then 
the observed system is completely time symmetric, and the underlying system is the continuous time version of the Brownian
clock discussed in the main text. To see this, we note that since $A$ and $C$ consist of single Markovian states,  as soon as
the system arrives at either $A$ or $C$, the future trajectory is independent of the past. Therefore, to prove time reversal
symmetry we need only have $\mathbb{P}(t,A\to B \to C) = \mathbb{P}(t,C\to B \to A)$, stating that the probability of seeing a transition
$A\to B \to C$, with waiting time $t$ in $B$ is the same as the time reversed trajectory, $C \to B \to A$ with wait time $t$ in $B$~\cite{SHorowitz_nocurrent}.
This is true if $\lambda = \mu$, and thus in this case the observed system is time symmetric, with relative entropies giving a trivial
bound on the entropy production rate. However, in this case, optimization finds the exact bound, Fig.~\ref{fig:AnalyticMin}B.

When the waiting time parameters are different, $\lambda\neq \mu$, the observed waiting time distributions may differ, and thus the relative entropy
of forward and reverse trajectories is non-zero. We compare the optimization bound to a bound that uses the information in the waiting time distributions~\cite{SHorowitz_nocurrent}.
In the the case of the example system, 
Fig.~\ref{fig:AnalyticMin}A, the waiting time bound becomes
\be
\sigma_{WTD} = n_{ABC} \int_{0}^{\infty} \psi(t|ABC) \log \left( \frac{\psi(t|ABC)}{\psi(t|CBA)} \right) 
+ n_{CBA} \int_{0}^{\infty} \psi(t|CBA) \log \left( \frac{\psi(t|CBA)}{\psi(t|ABC)} \right),
\ee
where $\psi(t|ABC)$ is the waiting time distribution at $B$ conditional on the path $A\to B\to C$,
and the overall bound is $\sigma_1 + \sigma_{WTD}$, ($\dot{S}_{aff} + \dot{S}_{WTD}$ in
Ref.~\cite{SHorowitz_nocurrent}). 
For the example from Fig.~\ref{fig:AnalyticMin}, 
\be
\psi(t|ABC) = \frac{a^2}{a^2 + b^2} \lambda e^{-\lambda t} + \frac{b^2}{a^2+b^2} \mu e^{-\mu t},
\ee
which, together with $n_{ABC} = (a^2 + b^2)/(a+b)$, allows us to calculate the estimate $\sigma_{WTD}$
as a function of $p$ and $\lambda$, $\mu$. We rescale time so that $\mu = 1$ and vary $\lambda > \mu$.
We see that as $\lambda \to \infty$, $\sigma_{WTD}$ approaches the true value of entropy production, 
Fig.~\ref{fig:AnalyticMin}, but as $\lambda \to \mu = 1$, the estimate approaches 0 (note $\sigma_1=0$ here). 
To obtain a reasonable bound for the waiting time estimator, a separation of time scales is needed, in addition to observing
enough conditional transitions to construct both waiting time distributions.
It is, in principle, possible to construct systems where $\sigma_{WTD} > \sigma_2$, although
this does not apply for any of the synthetic and experimental data considered in the present study. In any case, if sufficient
trajectories are observed to calculate waiting time distributions, one could define an estimator
$\hat{\sigma} = \max \{\sigma_2, \sigma_1 + \sigma_{WTD} \}$, to construct what is currently the best possible estimate.

\section{Switching random walker}
In Fig.~2 of the main text, we described the switching biased random walker to compare different entropy production rate estimators.
To calculate the estimates $\sigma_1$ and $\sigma_2$, we coarse-grained by making the 
system periodic modulo 3, so there were 3 observed macrostates. 
In this simple system we can calculate analytically the transition statistics, with
right steps occuring at a rate $\frac{1}{2} (q_1 + p_2)$, left steps at a
rate $\frac{1}{2} (p_1 + q_2)$. The exact entropy production rate is 
\be
\sigma_e = \frac{1}{2} \logfn{p_1}{q_1} + \frac{1}{2} \logfn{p_2}{q_2},
\ee
while the $\sigma_1$ estimate is 
\be
\sigma_1 = \frac{1}{2}\left[ p_2 + q_1 - p_1 - q_2 \right] \log \left( \frac{p_2 + q_1}{p_1 + q_2} \right) .
\ee
To find the $\sigma_2$ estimate, we take three neighboring observed states, 
and calculate $n_{I-1,I,I+1}$. To do so, we note that if $\ell_1$ is the probability of
eventually moving to $I+1$ given that the system is in the upper state of~$I$ in Fig.~2A of the main text, and $\ell_2$ the corresponding probability for the lower state, then
\begin{align}
\ell_1 = r \ell_2 + p_1, 
\qquad\qquad
\ell_2 = r \ell_1 + q_2, 
\notag
\end{align}
and therefore 
\be 
n_{I-1,I,I+1} = p_1 \ell_1 + q_1 \ell_2.
\ee
For generic $p_1 \neq p_2$ we input the calculated observables into the optimization program to find the entropy
estimate, for $p_1=p_2$, we can apply the analytic solution from the previous section.

For comparison, another method to estimate the entropy production rate is to use the thermodynamic uncertainty relation (TUR)
for inference, which states
\be 
\Sigma_\tau \geq 2 k_B \frac{\langle J_\tau \rangle^2}{\text{Var}(J_\tau)}
\ee
where $\Sigma_\tau$ is the increase in entropy in time $\tau$, and $J$ is any current. 
To estimate $\sigma = \Sigma_\tau/\tau$ (in steady state), we take an 
infinitesimal time $\tau$, and analytically calculate the current fluctuations one would see for an
ensemble of walkers prepared in the steady state distribution.
The natural current $J$ to take is the number of net right steps, so 
\begin{align}
\langle J_\tau \rangle   &= \tau \left( \frac{p_1+q_2}{2} - \frac{p_2 + q_1}{2} \right) + O(\tau^2)\\
\langle J_\tau^2 \rangle &= \tau \left( \frac{p_1+q_2}{2} + \frac{p_2 + q_1}{2} \right) + O(\tau^2),
\end{align}
so that 
\be
\sigma = \Sigma_\tau / \tau \geq k_B \frac{ \left[ (p_1 + q_2) - (p_2 + q_1) \right]^2}{p_1 + q_1 + p_2 + q_2},
\ee
providing another bound on the entropy production rate. The comparison in Fig.~2C-D of the main text show that the $\sigma_2$ bound improves on both the $\sigma_1$ and the TUR estimators, especially when net fluxes become small.

\section{Coarse graining in time}
A challenge with the current approach is that if we are observing an experimental system, we do not
get to see all of the transitions, but only see the state of the system at fixed time points due to
limited resolution. This means that instead of observing the Markovian process, we are really
observing a Markov chain. Here, we prove that by applying our method to the observed Markov chain
statistics, we still get a lower bound on the rate of entropy production, and hence limited experimental 
resolution poses no fundamental issue to our estimator.

Consider a Markov process, $X_t$ where the probability of transitioning from $i$ to $j$ after a time $t$ is
\be
\mathbb{P}(X_t = j | X_0 = i) = P_{ij}(t).
\ee
If we only observe this process at times $t = 0,T,2T,\dots$, we observe a Markov chain with transition
matrix $P_{ij}(T)$, and stationary distribution $\pi_i$ (the same stationary distribution as the Markov
process). This Markov chain produces entropy at a rate
\be
\sigma_{\text{MC}} = \frac{1}{T} \sum_{i\neq j} \pi_i P_{ij}(T) \log \left(
\frac{\pi_i P_{ij}(T)}{\pi_j P_{ji}(T)} \right).
\ee
Compare that to the rate at which entropy is produced by the Markov process over the time interval $[0,T]$,
\be
\sigma = \frac{1}{T} \int_{\text{paths}} \mathbb{P}(\text{forward path})
\log \left( \frac{\mathbb{P}(\text{forward path})}{\mathbb{P}(\text{reverse path})} \right).
\ee
For example a path could be $i \to j \to k$, with jumps at times $0 < r < s \leq T$. Consider
the function 
\be
g(x,y) = x \log \frac{x}{y},
\ee
which appears in the rate of entropy production. The corresponding Hessian
\be
Hg = \left( \begin{array}{cc} 
		\frac{1}{x} & - \frac{1}{y} \\
		-\frac{1}{y} & \frac{x}{y^2} \end{array} \right) ,
\ee
is positive semi-definite for $x,y > 0$. This implies convexity meaning,
\be 
x_1 \log \frac{x_1}{y_1} + x_2 \log \frac{x_2}{y_2} \geq (x_1 + x_2) \log \left( 
\frac{x_1 + x_2}{y_1 + y_2} \right),
\ee
and
\be
\int_{\vec{\lambda}} \; x(\vec{\lambda}) \log \frac{x(\vec{\lambda})}{y(\vec{\lambda})} 
\geq  \left[\int_{\vec{\lambda}} x(\vec{\lambda}) \right] \log 
\frac{\left[\int_{\vec{\lambda}}x(\vec{\lambda})\right]}{\left[\int_{\vec{\lambda}}y(\vec{\lambda})\right]},
\ee
for some functions of a vector of parameters $\vec{\lambda}$.
In particular, since 
\be
\int_{\text{paths}} = \sum_{i , j} \int_{\text{paths from } i \to j}, 
\ee
where a path from $i$ to $j$ means only that $X_0 = i$, $X_T = j$, then 
\be 
\sigma \geq \frac{1}{T} \sum_{i , j} \left[ \left(\int_{i \to j} \mathbb{P}(\text{forward}) \right)
\log \left( \frac{\int_{i \to j} \mathbb{P}(\text{forward})}{\int_{i \to j} 
	\mathbb{P}(\text{reverse})} \right) \right],
\ee
but 
\be
\pi_i P_{ij}(T) = \int_{i \to j} \mathbb{P}(\text{forward}),
\ee
and
\be
\pi_j P_{ji}(T) = \int_{j \to i} \mathbb{P}(\text{forward})
=\int_{i \to j} \mathbb{P}(\text{reverse}),
\ee
hence
\be
\sigma \geq \sigma_{\text{MC}}.
\ee
Therefore, by calculating the entropy produced by the discrete time Markov chain, we underestimate
the entropy produced by the true Markovian system. We can follow the same ideas as before, and minimize
over all Markov chains with the same first order and conditional statistics. However, given any
Markov chain with mass transfer per step $n_{ij}$, we can define a Markov process with the same mass
transfer per $T$, with the same rate of entropy production, so we can proceed as described before.

\section{Discretizing continuous time processes}
One can show that a Markov chain, under certain consistency conditions which can be enforced,
will converge to any given stochastic differential equation~\cite{SKloeden1992}. Indeed, numerically,
any time-discretized solution of a stochastic differential equation is really simulating a Markov
chain, since finite precision arithmetic implicity discretizes space as well. As shown in the
previous section, it does not matter to our estimator whether the underlying process is a Markov 
chain, or a Markovian process with the same values of $n_{ij}$ and hence the same rate of entropy 
production. Therefore, our method can be applied to any system described by a stochastic differential 
equation, provided that the rate of entropy production can be related to relative probabilities
of forward and reverse transitions, as in Eq.~\eqref{eq:Ent} above.

\section{Example: Brownian Motion}
Here we include an example, showing our framework can bound the entropy production for a 
model continuous system described by a Langevin equation. Consider Brownian motion with drift parameter $\mu$,
with trajectory described by the stochastic differential equation,
\be
\mathrm{d}X(t) = \mu \mathrm{d}t + \nu \mathrm{d} B(t),
\ee
where $dB(t)$ is an increment of a Weiner process~\cite{SKloeden1992}, and $\nu$ denotes the noise strength.
This could be a model for a Brownian clock, if $X(t)$ takes values in a period domain. 
Taking a discretization, with time step $\Delta t$, the probability of a forward path
$(X_0,\dots,X_N)$ is 
\be
\mathbb{P}_f = \mathbb{P}(X_0) \prod_{i=1}^{N} \frac{1}{\sqrt{2\pi \Delta t \nu^2}} e^{ - \frac{(X_i - X_{i-1} 
-\mu \Delta t)^2} {2 \Delta t \nu^2}, }
\ee
whereas the probability of observing the reverse path is 
\be
\mathbb{P}_r = \mathbb{P}(X_N) \prod_{i=1}^{N} \frac{1}{\sqrt{2\pi \Delta t \nu^2}} e^{ - \frac{(X_{i-1} - X_{i} 
-\mu \Delta t)^2} {2 \Delta t \nu^2}, }
\ee
hence
\be
\log \frac{\mathbb{P}_f}{\mathbb{P}_r} = \sum_{i=1}^N \frac{1}{2 \Delta t \nu^2} \left[
-(X_i - X_{i-1} - \mu \Delta t)^2 + (X_{i-1} - X_i -\mu \Delta t)^2 \right] = \frac{ 2\mu}{\nu^2} ( X_N - X_0),
\ee
and so
\be 
\left\langle \log \frac{\mathbb{P}_f}{\mathbb{P}_r} \right\rangle = \frac{2\mu^2}{\nu^2} (N\Delta t).
\ee
In the limit where $\Delta t \to 0$, $N\Delta t \to t$, the steady rate of entropy production in
this system becomes $\sigma = 2\mu^2/\nu^2$.

\begin{figure}\centering
\includegraphics{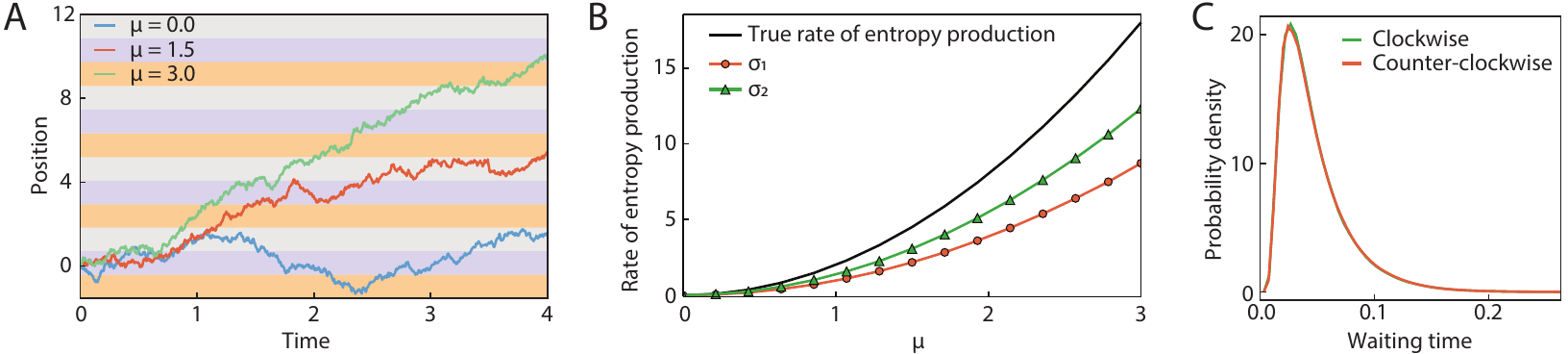}
\caption{\label{fig:BrownianClock} Bounding the rate of entropy production for Brownian motion.
(A) Sample trajectories for Brownian motion described by $\mathrm{d}X = \mu\mathrm{d}t + \nu \mathrm{d}B$,
for $\nu= 1$ and various values of $\mu$. To bound the entropy production we divide space into 3 regions
repeated periodically (purple,grey,orange). (B) Bounds on the rate of entropy production for
the $\sigma_1$ estimator (orange), and the $\sigma_2$ estimator (green), with the true rate of 
entropy production $\sigma = 2\mu^2/\nu^2$ (black). (C) Comparison of waiting time distributions 
given two consecutive clockwise or counter clockwise transitions with regions of width 1/3, and 
with parameters $\mu = 2$, $\sigma = 1$, showing the forward and reverse waiting time distributions 
are identical. }
\end{figure}

Given a trajectory observed at finite time resolution we can coarse grain space into
$3M$ regions, corresponding to 3 macrostates, Fig~\ref{fig:BrownianClock}A. We further optimize over $M$ to find the 
optimal bound, given the temporal resolution. We find that the $\sigma_1$ estimator finds a reasonable bound,
with the $\sigma_2$ estimator finding a bound around $20\%$ better, Fig.~\ref{fig:BrownianClock}B. In this system,
which has observable currents, $\sigma_1$ and $\sigma_2$ perform similarly, but as seen in the bacterial flagella
motor example, if observable currents are absent, $\sigma_2$ will perform significantly better than $\sigma_1$.
The waiting time distributions appear to be the same for the forwards and backwards paths, so 
no additional information is gained from the waiting time estimator, Fig.~\ref{fig:BrownianClock}C.

\end{document}